\documentclass{aa}  
\usepackage{graphicx}
\usepackage{txfonts}
\usepackage{amsmath}
\usepackage{amssymb}
\usepackage{multicol}
\usepackage{natbib}
\usepackage{comment}
\usepackage[colorlinks=true,allcolors=blue]{hyperref}
\newcommand{\ergs}[0]{\rm erg\,\rm s^{-1}}

\newcommand{\gcm}[0]{\text{g}{\,}\text{cm}^{-3}}

\begin{document} 
   \title{Relativistic hydrodynamics simulations of supernova explosions within extragalactic jets}

    \titlerunning{Relativistic hydrodynamics simulations of supernova explosions within extragalactic jets}
   
   \author{B. Longo
          \inst{1}, M. Perucho \inst{1,2}, V. Bosch-Ramon \inst{3}, 
          J.M. Mart\'i \inst{1,2},
          G. Fichet de Clairfontaine \inst{1} 
          }

    \authorrunning{Longo et al.}
          
   \institute{ Departament d'Astronomia i Astrof\'isica, Universitat de València, C/ Dr. Moliner,
    50, 46100, Burjassot, Val\`encia, Spain.
         \and
             Observatori Astron\`omic, Universitat de València, C/ Catedràtic José Beltr\'an 2, 46980, Paterna, Val\`encia, Spain.
          \and
             Departament de F\`isica Quàntica i Astrof\'isica, Institut de Ciències del Cosmos (ICC), Universitat de Barcelona (IEEC-UB), Mart\'i i Franquès 1, E08028 Barcelona, Spain.
             }
   \date{Accepted for publication in A\&A}
  \abstract 
   {Jets in active galactic nuclei have to cross significant distances within their host galaxies, meeting large numbers of stars of different masses and evolution stages in their paths. Given enough time,
   supernova explosions within the jet will eventually happen, and may have a strong impact on its dynamics, potentially triggering powerful non-thermal activity.}
   {We carried out a detailed numerical study to explore the dynamics of the interaction between the ejecta of a supernova explosion and a relativistic extragalactic jet.}
   {By means of relativistic hydrodynamics simulations using the code \texttt{RATPENAT}, we simulated the jet-ejecta interaction in two different geometries or scenarios: a two-dimensional, axisymmetric simulation, and a three-dimensional one, which includes the orbital velocity of the exploding star. In both scenarios, the supernova ejecta is located within the jet at a distance of $\sim 1$~kpc from the central black hole, which is the spatial scale at which these events are most likely.}
   {Although initially filling a region much smaller than the jet radius, the ejecta expands and eventually covers most of the jet cross section. The expansion is enhanced as more energy from the jet is converted into kinetic and internal energy of the ejecta, which also favors the ejecta disruption, all this occurring on timescales $\sim 10^{4}\,$yr. Although a complete numerical convergence of the results is unattainable given the subsonic, turbulent nature of the interaction region, the simulations are consistent in their description of the gross morphological and dynamical properties of the interaction process.}
   %The simulations show convergence with resolution.}
   {At the end of the simulations, the supernova ejecta has already partially mixed with the relativistic jet. The results also suggest that the jet-ejecta interaction may be a non-negligible non-thermal emitter. Moreover, due to efficient mixing, the interaction region can be a potential source of ultra-high-energy cosmic rays of heavy composition.}
   \maketitle

%-------------------------------------------------------------------

\section{Introduction}

Being launched from the surroundings of a supermassive black hole (SMBH) at the center of an active galactic nucleus (AGN), relativistic extragalactic jets are collimated outflows that propagate through the galactic medium within their first kiloparsecs \citep[][]{1984RvMP...56..255B,2019ARA&A..57..467B}. The inner galactic regions are known to contain large amounts of gas and stars and related objects (e.g., \citealt{burbidge70}; \citealt{kauffmann03}; \citealt{hub06}; \citealt{shao10}). The interaction between the jet and such objects is unavoidable and has been well studied, with the focus either on dynamical \citep[e.g.,][]{komissarov94,bosch12,perucho14,perucho17,castillo2021} or radiative effects \citep[e.g.,][]{barkov10,bosch12,araudo13,delacita16,2025A&A...693A.270F}. Very close to the SMBH, it has been proposed that this kind of interaction and its resulting emission often occur as short and relatively frequent events (e.g., \citealt{aha17}; \citealt{del19}; \citealt{araudo20}; \citealt{kurfurst24}), while further downstream of the jet the interaction timescales can be much longer and several events may coexist \citep[e.g.,][]{2015MNRAS.447.1001W,torres-alba19}. Unless the obstacle rest mass is large or the jet power is low, gas clouds and stellar wind envelopes that interact with the jets can be completely disrupted \citep{bosch12}, in which case the mass of the object determines the maximum energy dissipated by the interaction, and the jet spatial scale characterizes the overall evolution time \citep[e.g.,][]{khangulyan13}.

Among the object types that can face extragalactic jets, the ejecta produced by supernova (SN) explosions is perhaps one of the least studied ones \citep[e.g.,][]{bla79,fed96,bed99}. However, core-collapse SN explosions occurring inside jets in galaxies with non-negligible star-formation can lead to significant non-thermal emission, as proposed, e.g., in \cite{vieyro19}. A similar situation regardless of the galaxy star-formation rate can also be realized when type~Ia (thermonuclear) SN explosions interact with the jet \citep{tor19}. Although the occurrence of these events is low,\footnote{It is proportional to the galactic volume filled by the jet, which is roughly proportional to the square of the jet-to-galaxy radius ratio.} the duration of such an interaction is long\footnote{The duration timescale is approximately given by the jet distance from the SMBH to the interaction site over $c$.}. Therefore, given that most of SN explosions occur within a few kpc from the center of galaxies, \cite{bosch23} concluded that up to $\sim 10$\% of jetted AGN may host a jet-SN interaction at some stage of its evolution. Since no strong qualitative differences are expected between core-collapse and thermonuclear supernovae (SNe), both star- and non-star-forming jetted AGN are worth being considered when studying these events, with the most likely height scale (jet length from the SMBH to the interaction site) between hundreds and one thousand parsecs \citep[see][and references therein]{bosch23}. 

Initially, the SN ejecta expands freely until its expansion is stopped in the upstream jet direction when the ejecta and jet ram pressures balance each other. At this point, the thermodynamical properties and dynamics of the ejecta become dominated by the jet impact. The expansion rate of the shocked gas is then expected to accelerate, and previous work on jet-cloud interactions suggests that the shocked ejecta may get disrupted and mixed with the jet flow \citep[see][and references therein]{bosch23}. For instace, this was suggested by two-dimensional (2D) axisymmetric simulations presented in \citet{vieyro19}, but the simplified geometry and low resolution of the simulations prevented arriving at strong conclusions. Moreover, the fact that the star is orbiting the center of the galaxy can lead to some anisotropy in the jet-SN interaction even if the associated velocity is small with respect to that of the jet, which adds to the limitations of 2D simulations. More detailed and realistic simulations are also needed to better understand non-thermal processes in these interactions.

Motivated by the lack of thorough dynamical studies of extragalactic jet-SN ejecta interactions and as a first step, we investigated this scenario using numerical simulations in the context of relativistic hydrodynamics (RHD). Simulations including a magnetic field will be presented in future work. Two different cases were considered here: As a first exploration and for computational time purposes, we first ran 2D axisymmetric RHD simulations, followed by three-dimensional (3D) runs. 
We adopted two different numerical resolutions to study the interaction scenario 1) with the largest accuracy attainable with affordable supercomputing resources, and 2) in the largest spatial and temporal domains, without compromising the numerical consistency of the results.
%We adopted different resolutions to analyze the convergence of the results. 
The paper is structured as follows: in Sect.~\ref{sce}, we present the physical scenario in more detail, and in Sect.~\ref{simulations_ref} we describe the 2D and 3D RHD simulation runs for this study. In Sect.~\ref{resu}, we present the results obtained for the different cases considered, and finally in Sect.~\ref{disc} we discuss our results and give our conclusions. 

\section{Jet and supernova ejecta properties}\label{sce}

We consider that the interaction occurs at a distance of $1$~kpc from the SMBH, when the jet is completely developed, its radius is $\sim 100$~pc, and the stellar density is still large. For simplicity, we assumed an opening angle of $0.2$~radian, which implies a jet radius of $R_{\rm j} = 100$~pc. The simulations are set up at a point in which the SN is large enough to be properly resolved in terms of cell numbers per diameter. Our set-ups are, nevertheless, spatially scalable in the sense that the jet radius determines the jet ram pressure, so that adopting a narrower (wider) jet would imply the interaction being located proportionally farther (closer) to the SMBH. The simulations can also be scaled to other jet-ejecta interaction scenarios in which, ceteris paribus, the jet ram pressure-to-obstacle mass ratio is the same. Thermal cooling may break this symmetry, but this effect can be neglected given the involved low densities, high temperatures and relatively short evolution times at the considered jet scales. 
%Furthermore,  they can be scaled to other jet-obstacle scenarios \citep[see in this regard the $D$ parameter given in eq.~6 in][]{khangulyan13}. 
Finally, following \cite{bosch23}, we adopted an intermediate jet power within typical AGN jet values, $L_{\rm j}=10^{44}\,\ergs$.
%\footnote{
%We have used $L_{\rm j}=\rho(hW - 1)W v A$, where $A$ is the jet cross-section. 
%Here we exclude the rest-mass density.} 
\footnote{Here $L_{\rm j}$ refers to kinetic power, i.e., total power excluding rest-mass energy flux.}
On the one hand, this power is high enough to make plausible the acceleration of ultra-high-energy cosmic rays (UHECR) at the jet-SN interaction region, and, on the other hand, low enough to increase the odds of these events happening in the local Universe \citep{bosch23}. 

The SN ejecta initial conditions have been characterized in a simplified manner, given the difference in spatial scales involved, to avoid an excessive computational demand. Therefore, we modeled the initial SN state as a uniform highly pressured and dense bubble at rest, with a total internal energy similar to the typical kinetic energy of an SN ejecta, $E_{\rm SN}=10^{51}\,\rm erg$ \citep{leahy17}, and a mass of $M_{\rm SN}=2M_\odot$, between those released by thermonuclear (SN~Ia) and core-collapse SNe. The initial ejecta density is thus: 
\begin{equation}
\rho_{\rm SN} = \frac{M_{\rm SN}}{\frac{4}{3}\pi R_{\rm SN}^{3}},
\end{equation}
where $R_{\rm SN}$ is the initial SN ejecta radius, which is fixed in our simulations to a given initial value. The value of $R_{\rm SN}$ depends on the expected maximum upstream expansion of the ejecta as measured from the SN initial location. Under these conditions, the initial expansion speed can be estimated as $\sim\sqrt{E_{\rm SN}/M_{\rm SN}}\sim10^{4}\,\rm km\,\rm s^{-1}$. The ejecta will expand at this speed in all directions until the jet ram pressure becomes significant and upstream expansion decelerates.

From the very beginning, the jet-ejecta interaction develops a bow-shaped structure made of shocked jet flow common in jet-obstacle interactions \citep[e.g.,][]{komissarov94,bosch12}. The expected maximum expansion point in the jet upstream direction is located at a distance from the SN explosion origin that can be approximated by imposing pressure balance between the shocked jet and ejecta:
\begin{equation} \label{eq:max}
R_{\rm max} = \left(\frac{5\,E_{\rm SN}}{2\pi p_{\rm ram,j}}\right)^{1/3},    
\end{equation}
where $p_{\rm ram,j}$ is the jet ram pressure:
%\begin{align}
%    p_{\rm ram,j}=v_{\rm j}\frac{L_{\rm j,\rm t}}{ \pi R_{\rm j}^{2} c^2}\,,
%\end{align}
%where $L_{\rm j,\rm t}$ and $R_{\rm j}$ are the jet total kinetic power\footnote{Here we include the rest-mass density.} 
%component$L_{\rm j}=\rho\,h\,W^2\,v\,A$ in this case) 
%and radius, respectively.
\begin{align}
    p_{\rm ram,j}=\frac{v_{\rm j}L_{\rm j}W_jh_{\rm j}}{ \pi R_{\rm j}^{2} c^2(W_{\rm j}h_{\rm j}-c^2)}+p_{\rm j}\, ,
\end{align}
where $p_{\rm j}$ is the jet pressure, $v_{\rm j}$ the jet flow speed, $W_{\rm j}$ and $h_{\rm j}$ the jet Lorentz factor and specific enthalpy, respectively (where $h_{\rm j}=c^2+\varepsilon_{\rm j}+p_{\rm j}/\rho_{\rm j}$, with $\varepsilon_{\rm j}=p_{\rm j}/(\Gamma-1)$ being the specific internal energy and $\Gamma$ the adiabatic index). Beyond this moment, the ejecta evolution becomes dominated by the jet ram pressure. The initial ejecta radius in the simulation is taken to be a small fraction of $R_{\rm max}$, that is, $R_{\rm SN}=f_{\rm R}\,R_{\rm max}$ with $f_{\rm R}\ll 1$, to allow us to start following the evolution of the ejecta while it is still well within its free expansion phase.

The Lorentz factor of the jet is taken as $W_{\rm j}=2$ ($v_{\rm j} \approx 0.866$ $c$) and the specific enthalpy as $h_{\rm j}=1.1\,c^{2}$, so that the unshocked jet density and gas pressure can be derived from
\begin{align}
    \rho_{\rm j} =
    %\;& 
    \frac{L_{\rm j}}{\pi R_{\rm j}^{2}v_{\rm j}W_{\rm j}(W_{\rm j}h_{\rm j}-c^{2})} %\\
%    \approx \;& 6\times10^{-30} \gcm \left(\frac{L_{\rm j}}{10^{44}\,\ergs}\right)\left(\frac{R_{\rm j}}{100 \text{pc}}\right)^{-2}\left(\frac{v_{\rm j}}{0.866c}\right)^{-1}
, 
%\nonumber\\
%    p_{\rm j} = \;& (h_{\rm j}/c^{2}-1)\rho_{\rm j}\,c^{2}(\Gamma-1)/\Gamma\,.
\quad p_{\rm j} = \frac{\Gamma-1}{\Gamma}(h_{\rm j}/c^{2}-1)\rho_{\rm j}\,c^{2}\,,
\end{align}
%With the jet ram pressure and the kinetic energy of the SN explosion, one can obtain the stagnation point:
%\begin{align}
%R_{\rm eq} \approx 11\;\text{pc} \left(\frac{p_{\rm ram,j}}{10^{-8}\,\ergcm}\right)^{-1/3}\left(\frac{E_{\rm SN}}{10^{51}\,\text{erg}}\right)^{1/3}.
%\end{align}
to give $\rho_{\rm j} \approx 6\times10^{-30} \gcm$, $p_{\rm j} \approx 2\times10^{-10} \rm dyn\,\rm cm^{-2}$.
Finally, with the jet flow velocity fixed, the jet ram pressure and the maximum expansion size of the SN ejecta can be established: $p_{\rm ram, \, j} \approx 1.8\times 10^{-8}$ erg cm$^{-3}$, $R_{\rm max} \approx 11$ pc. Then, at an expansion speed of $\sim 10^4$~km~s$^{-1}$, the ejecta would reach this point after $\sim 10^3$~yr.

Both in 2D and 3D, two different initial SN radii are simulated: $10$ (S1) and $5$ (S2) times smaller than the maximum expansion radius (i.e., $f_{\rm R}=0.1, 0.2$); thus we have $R_{\rm SN}=1.1,2.2\;\text{pc}$. The number of cells per initial SN radius is the same in both simulations, so the effective spatial resolution is doubled in S1 with respect to S2. Furthermore, we introduced an ejecta velocity of $200\,\rm km\,\rm s^{-1}$ perpendicular to the jet, mimicking the orbital velocity of the progenitor star and therefore introducing an asymmetry in the system. 

Despite the fact that the ejecta initial expansion speed is 
%$\sqrt{E_{\rm SN}/M_{\rm SN}}\sim 10^{4}\,\rm km\,\rm s^{-1}$, so 
$\sim 50$ times larger than the orbital one, the global structure of the shocked ejecta can develop
large asymmetries as it evolves. The grid surrounding the initial SN bubble was filled with a homogeneous jet flow.

Regarding the gas composition, we assumed, for both the SN ejecta and the jet, a neutral electron-proton gas with a leptonic mass fraction
$X_{\rm e} = m_{\rm e}/m_{\rm p} \approx 5.5\times10^{-4}$,
yielding an effective mass per particle $m_{\rm eff}\approx m_{\rm p}/2$. This is a reasonable plasma prescription if the jet has been significantly mass-loaded within the inner regions of the galaxy, which is expected \citep{perucho14,castillo2021}. The initial SN ejecta density is determined by $R_{\rm SN}$, $1.1$~pc in S1 and $2.2$~pc in S2, resulting in $\rho_{\rm SN}= 2.4\times10^{-23}\gcm$, and $3\times10^{-24}\gcm$, respectively. Finally, the gas temperature is computed using 
%$T=p_{\rm SN}/\rho m_{\rm eff} k_b$, with $p_{\rm SN}$ being the initial ejecta pressure and is derived from $E_{\rm SN}$ and $R_{\rm SN}$. 
$T=m_{\rm eff} \, p/k_{\rm B} \,\rho$ for both jet and ejecta (with the initial ejecta pressure derived from $E_{\rm SN}$ and $R_{\rm SN}$). Thus, the ejecta temperature is $T_{\rm SN} \approx 10^{9}\,\text{K}$ at the start of the simulation; in the unshocked jet, the temperature is to $T_{\rm j} \approx 2\times10^{11}\text{K}$.

\section{Simulations}\label{simulations_ref}

To run the simulations, we used a finite-volume code, which solves the RHD equations in conservative form by means of high-resolution shock-capturing methods (HRSC). We used its OpenMP version in cylindrical coordinates for the axisymmetric 2D simulations, while we used the hybrid MPI + OpenMP version \citep[\texttt{RATPENAT};][]{perucho10}  for the 3D simulations in Cartesian coordinates.

In 3D Cartesian coordinates, the conservation equations in $c=1$ units can be written as follows:
\begin{align}
\frac{\partial \mathbf{U}}{\partial t} + \frac{\partial \mathbf{F^{i}}}{\partial x^{i}} = 0\,,
\end{align}
where the state vector $\mathbf{U}$ is:
\begin{align}
\mathbf{U} = (D,D_{\rm e},S^{j},\tau)^{T},
\end{align}
and the vectors of fluxes $\mathbf{F^i}$ is:
\begin{align}
\mathbf{F^{i}} = (Dv^{i},D_{\rm e}v^{i},S^{j}v^{i}+p\delta^{ij},S^{i}-Dv^{i})^{T}, 
\end{align}
with $i,j=1,2,3$ and summation over repeated indices implicit. The variables $D$, $D_{\rm e}$, $S^{i}$, and $\tau$ are, respectively, the total and leptonic rest-mass densities, the momentum density in each spatial direction, and the energy density (without the rest-mass energy density), defined in the laboratory frame, and are related to the quantities in the local rest-frame of the fluid as:
\begin{align}
D = \rho W, \;\; D_{\rm e} = \rho_{\rm e} W, \;\; S^{i} = \rho h W^{2} v^{i},\;\; \tau = \rho hW^{2} - p - D\,,
\end{align}
where $\rho$ and $\rho_{\rm e}$ are the total and leptonic densities, $v^{i}$ the components of the velocity of the fluid, $W$ the Lorentz factor ($W=(1-v^{i}v_{i})^{-1/2}$), $p$ the gas pressure, and the specific enthalpy is defined in $c=1$ units 
as:
\begin{align}
h = 1 + \varepsilon + p/\rho\,.
\end{align}
%with $\varepsilon$ the specific internal energy. 
The code uses the Synge equation of state \citep{synge57}, which includes protons and electrons as a mixture of relativistic Boltzmann gases. Finally, we use a tracer, $f$, which gives the jet-mass fraction, and allows us to trace the mixing between jet ($f=1$) and SN ejecta ($f=0$) materials.

%%%%%%%%%%%%%%%%%%%%%%%%%%%%%%%%%
%
\begin{figure}[!ht]
\centering
\includegraphics[width=\linewidth]{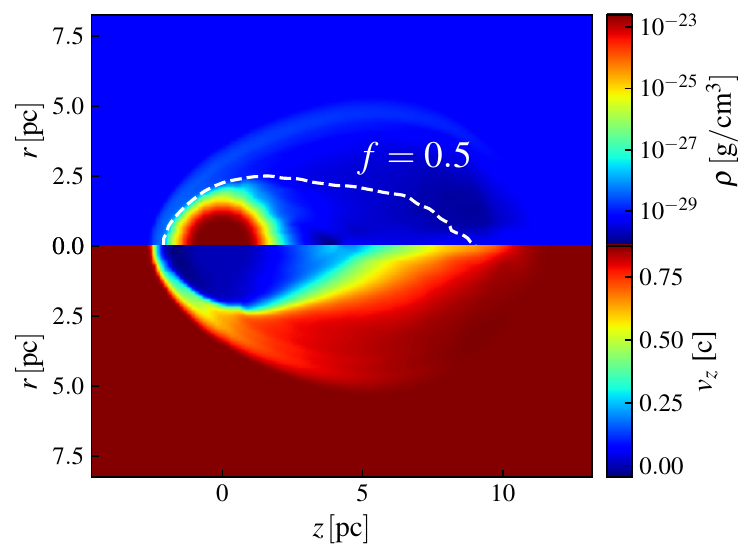}
\caption{Snapshot of a jet-SN interaction during the initial phase to illustrate the set-up of the 2D axisymmetric simulations. Upper half: rest-mass density, $\rho$. Lower half: axial velocity, $v_z$. The white dashed-line gives the jet-mass fraction contour for the tracer value $f=0.5$. The jet flow is filling the grid but in the ejecta region, and propagates from left to right.}
\label{2d_2msun_hr_rho_velz}
\end{figure}
%
%%%%%%%%%%%%%%%%%%%%%%%%%%%%%%%%%%

\subsection{2D axisymmetric simulations}

We first ran 2D axisymmetric simulations in which we set up a uniformly highly pressured and dense bubble of gas, which corresponds to the SN ejecta, surrounded by a homogeneous relativistic jet flow moving in the $Z$-direction, similarly as implemented in previous works \citep[e.g.,][]{vieyro19}.
%Taking into account that the jet radius is much larger than $R_{\rm SN}$, we limited our numerical experiment to the interior of the jet, so excepting the SN ejecta region, the jet fills the rest of the grid, as shown in Fig.~\ref{2d_2msun_hr_rho_velz}, and its streamlines are parallel.
Figure~\ref{2d_2msun_hr_rho_velz} zooms into the region around the SN ejecta during the free expansion phase. The figure shows a still spherical core of the remnant of the shocked ejecta and the development of a downstream-elongated tail. Upstream, a bow shock separates the region of interaction from the undisturbed homogeneous jet flow.

In these two-dimensional simulations, we used cylindrical coordinates and assumed axisymmetry. Accordingly, the left and right boundaries of the $Z$-axis are defined with an inflow and outflow condition, respectively, whereas in the radial direction, the conditions are reflection on the symmetry axis and outflow at the outermost boundary. 

In the first of our simulations, S1 ($R_{\rm SN}=1.1$~pc), the dimensions of the grid are $L_r\times L_z=96\;R_{\rm SN}\times192\;R_{\rm SN}$, with $768\times1536$ cells, resulting in a resolution of 8 cells/$R_{\rm SN}$. In the case of S2 ($R_{\rm SN}=2.2$~pc), the dimensions of the grid are $L_r\times L_z=80\;R_{\rm SN}\times120\;R_{\rm SN}$, with $640\times960$ cells, and the same number of cells per $R_{\rm SN}$. The ejecta is initially centered on the axis, at one fourth of the grid length (i.e., $z=L_z/4$). We ran the simulations on 32 cores in the supercomputing facility \textit{Llu\'{\i}s Vives} at Universitat de Val\`encia.

\subsection{3D simulations}

%%%%%%%%%%%%%%%%%%%%%%%%%%%%%%%
%
\begin{figure*}
\centering
\includegraphics[width=0.3807\linewidth]{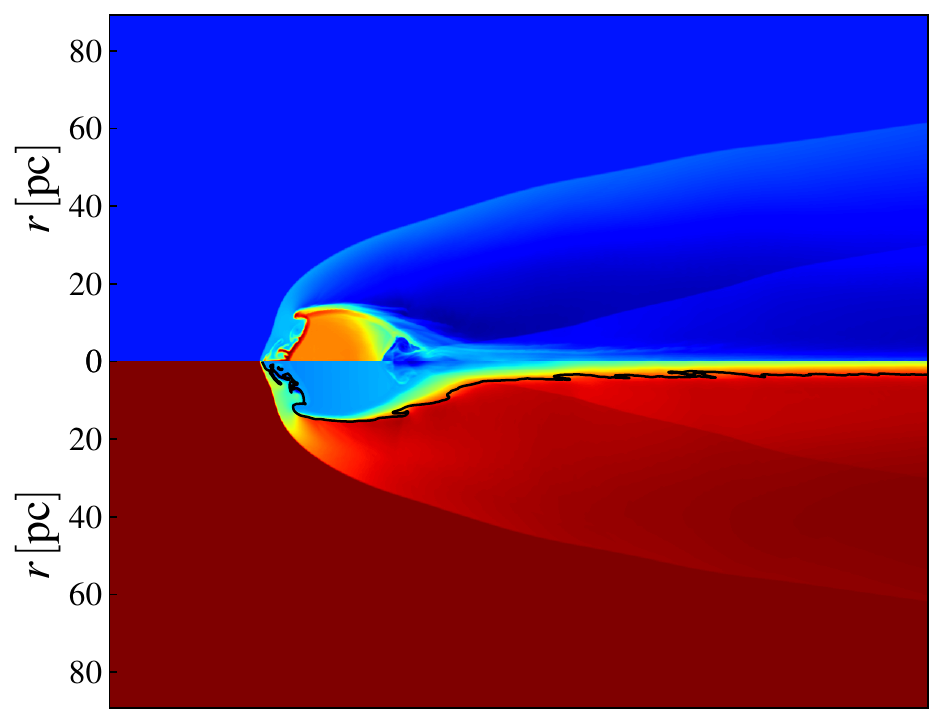} 
\includegraphics[width=0.42363\linewidth]{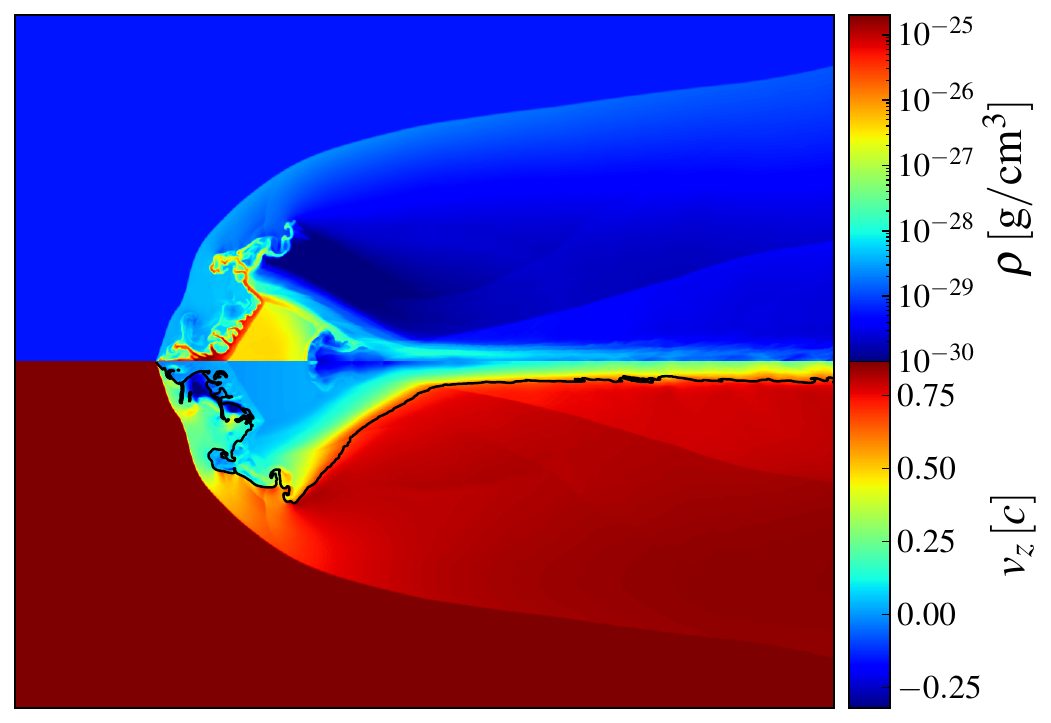}
\includegraphics[width=0.3807\linewidth]{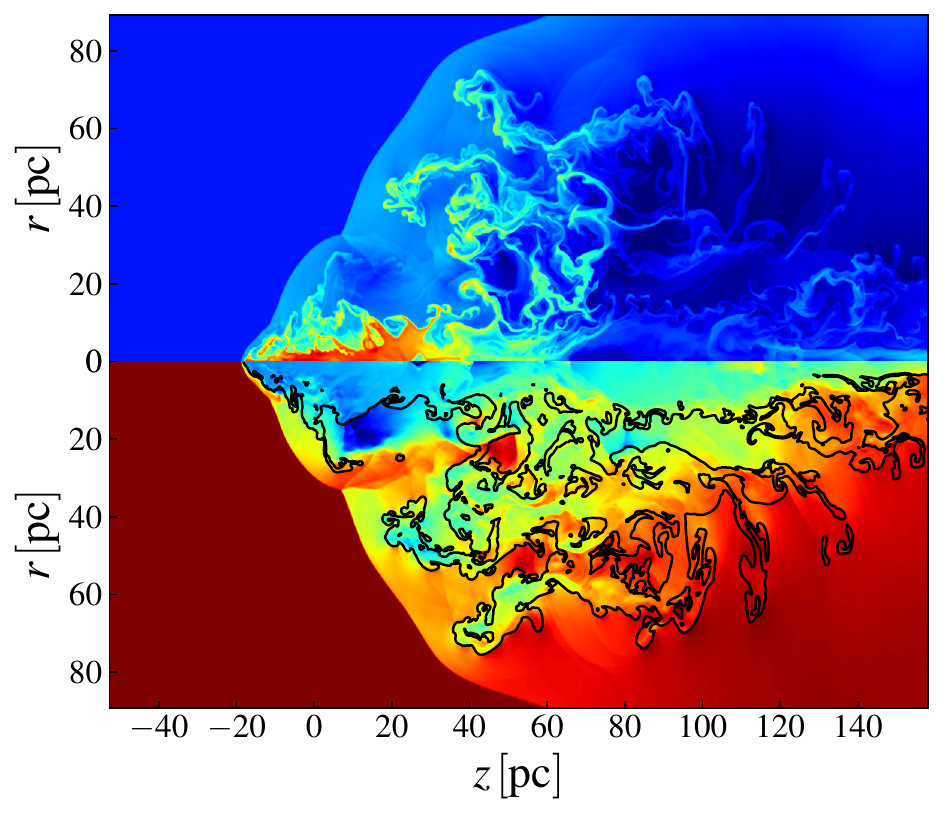} 
\includegraphics[width=0.42363\linewidth]{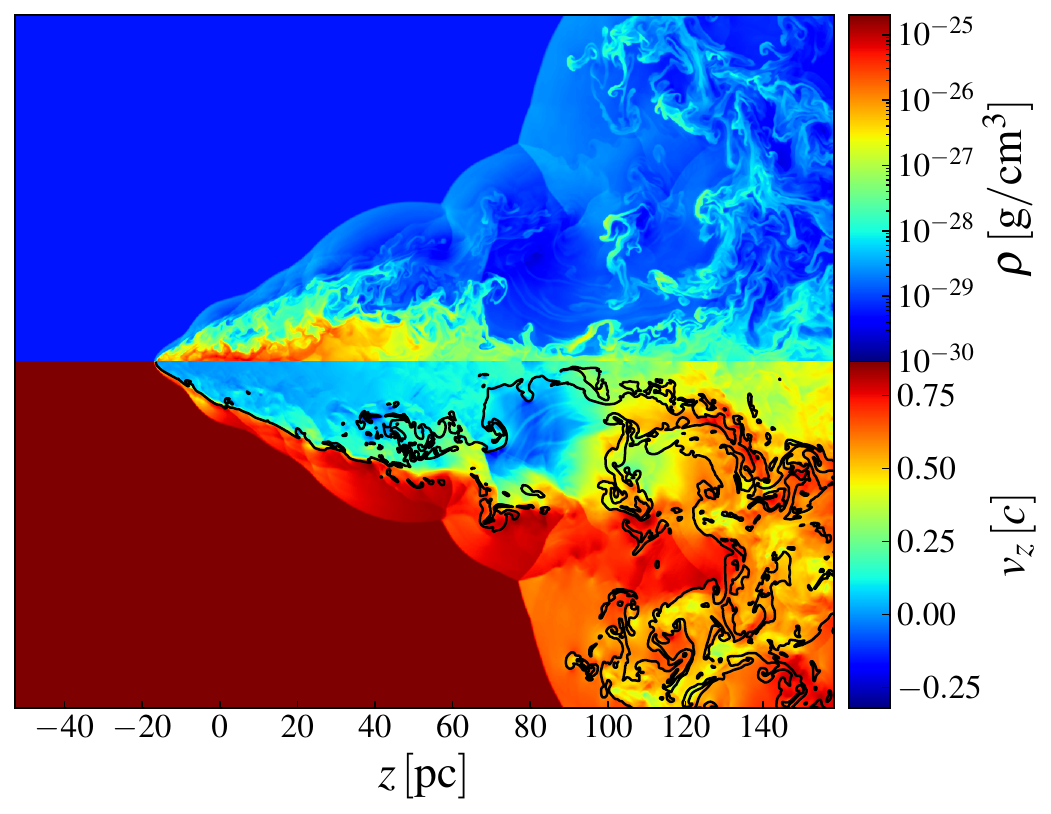}
\caption{Time evolution of the rest-mass density (upper half panels) and axial velocity (lower half panels, mirror of the upper panels) for S1 in 2D. The black lines give the contour of the jet-mass fraction for the tracer value $f=0.5$. The jet is propagating from the left to the right. Top left: $t\approx 1600$~yr from the start of the simulation; top right: $t\approx2300$~yr; bottom left: $t\approx3000$~yr; bottom right: $t\approx 3800$~yr.}
\label{2d_timeframe_hr}
\end{figure*}
%
%%%%%%%%%%%%%%%%%%%%%%%%%%%%%%%%%%%%

The physical set-up is very similar to the one we used in our 2D simulations, including numerical resolution. As a main difference, we included the orbital velocity ($v^x_{\rm orb} = 200$~km$\,$s$^{-1}$) of the progenitor star around the galactic center. The geometry of the grid is now Cartesian and is defined with a volume ($X,Y,Z$) $[-50,50]\,R_{\rm SN} \times [0,100]\,R_{\rm SN} \times [-50,50]\,R_{\rm SN}$ (i.e., $L_x=L_y=L_z=100\,R_{\rm SN}$) 
and $800^{3}$ cells for simulation S1 ($R_{\rm SN}=1.1$~pc), and ($X,Y,Z$) $[-40,40]\,R_{\rm SN} \times [0,80]\,R_{\rm SN}\times[-40,40]\,R_{\rm SN}$ (i.e., $L_x=L_y=L_z=80\,R_{\rm SN}$) 
and $640^{3}$ cells for S2 ($R_{\rm SN}=2.2$~pc). Those numbers result in a resolution of 8 cells/$R_{\rm SN}$ in both cases, the same as that used in the 2D simulations. We also ran the same 3D simulations, but for an initial ejecta at rest, to calibrate the effect of the initial ejecta motion perpendicular to the jet; the results of this simulation are shown in the appendix \ref{3dwm}.

%As in the 2D case, we defined the jet flow as filling the rest of the grid and propagating along the $Y$-axis. The initial SN ejecta is centered in the $XZ$-plane and at a distance $L_y/5$ from the left boundary along the $y$ direction in S1, and $L_y/8$ in S2.  
The initial SN ejecta is centered in the $XZ$-plane and at a distance $L_y/5$ from the left boundary along the $Y$-axis in S1, and $L_y/8$ in S2. The jet propagates along the $Y$-axis and fills the rest of the grid. Except for the left boundary of the $Y$-axis, where we inject the jet flow, all other boundaries of the numerical box are set with an outflow condition. The 3D simulations were also run on \textit{Llu\'{\i}s Vives}, using either 128 or 256 cores. Table \ref{table:1} in the Appendix shows the parameter values adopted for the simulations.

\section{Results}\label{resu}
\subsection{Simulation S1}

This simulation focuses on the evolution of a SN explosion, in which an ejecta of $2 \, M_\odot$ and $10^{51}$~erg initially fills a spherical region with radius $1.1$ pc ($\sim 0.1 \,R_{\rm max}$). In the following sections, we first present the results for the 2D axisymmetric simulation, and then those obtained for the 3D one.

\subsubsection{2D axisymmetric case}

%%%%%%%%%%%%%%%%%%%%%%%%%%%%%%%%%%%
%
\begin{figure*}
\centering
\includegraphics[width=.9\linewidth]{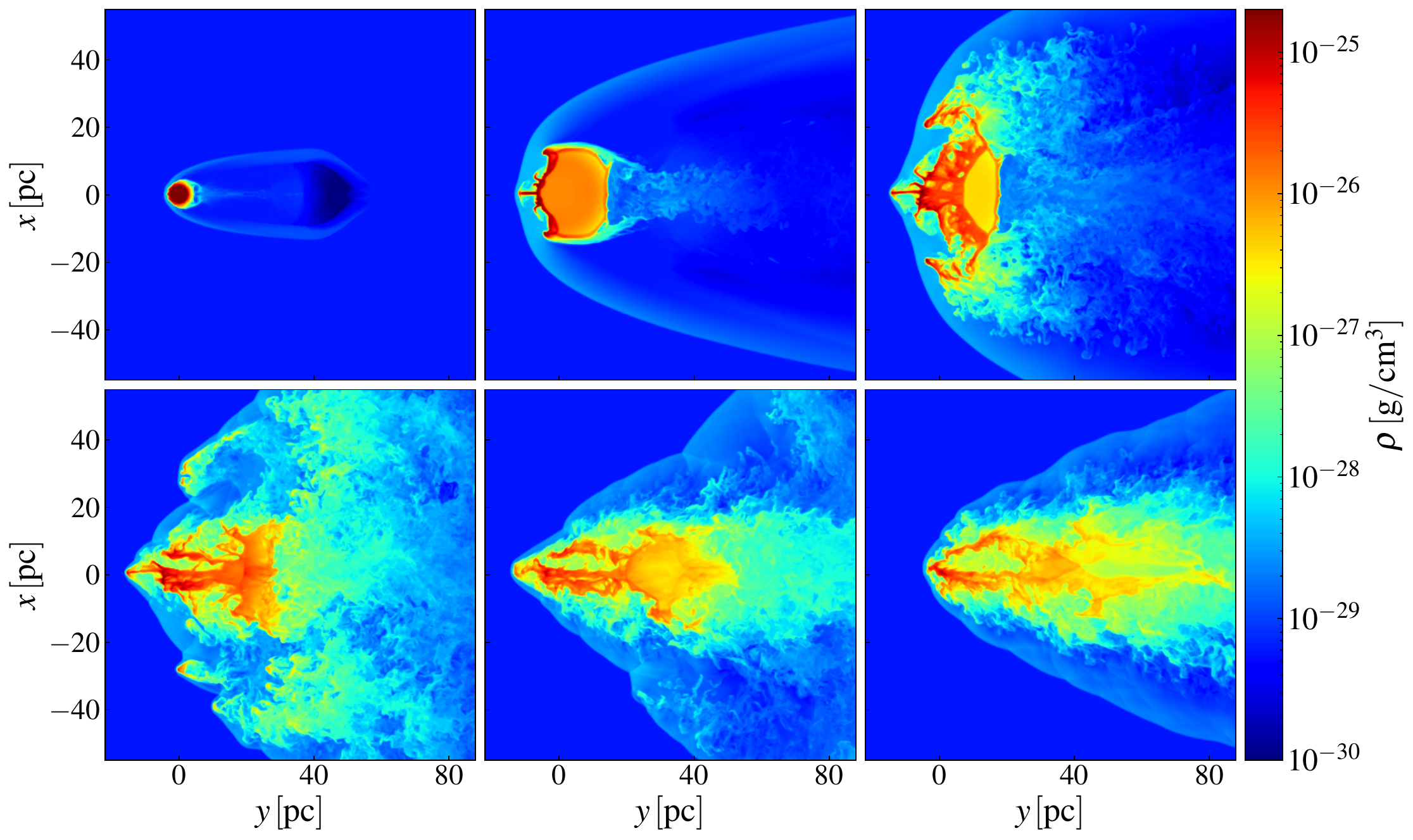}
\caption{Time evolution of the rest-mass density for S1 in 3D, with six 2D cuts in the $XY$-plane at $z=0$. The jet is propagating from the left to the right. Top left: $t\approx180$~yr; top middle: $t\approx1600$~yr; top right: $t\approx2300$~yr; bottom left: $t\approx3000$~yr; bottom middle: $t\approx3800$~yr; bottom right: $t\approx 4700$~yr.}
\label{3d_timeframe_rho_hr}
\end{figure*}
%
%%%%%%%%%%%%%%%%%%%%%%%%%%%%%%%%%%%%%%%

We show the evolution of the ejecta developing within the jet flow in Fig.~\ref{2d_timeframe_hr}, where the rest-mass density and axial flow velocities are shown in the upper and lower halves, respectively. A solid black line in the velocity plots indicates the jet-mass fraction at $0.5$. 
%The top left panel shows a snapshot after $1000\,\rm yr$ of evolution, about the end of the free expansion phase of the ejecta, when it has already expanded significantly and the shock in the jet is fully developed, forming an interaction region with radius $\sim 15$~pc. This is similar to $R_{\rm eq}$, as expected from the analytical calculation. The hot shocked jet material is being compressed at the shock interface, reaching a density of $\sim10^{-27}\,\rm g\,\rm cm^{-3}$. At this point, the ejecta material is partially shocked, and the unshocked region still keeps to a large extent its initial homogeneous structure while still freely expanding.
The top left panel shows a snapshot after $1600\,\rm yr$ of evolution, about the end of the free expansion phase of the ejecta, when it has expanded to a radius close to the maximum expected one, ${R_{\rm max}}$, as given by the analytical calculation (Eq.~\ref{eq:max}). At the same time, the bow shock in the jet is fully developed and the jet flow is heated and compressed through the discontinuity (up to a density of $\sim 4\times10^{-29}\,\rm g\,\rm cm^{-3}$). A backward shock (in the ejecta reference frame) detaches from the jet/ejecta impact point and starts to cross the (otherwise still homogeneous) ejecta (red/orange transition). Downstream, the tail formed with ejecta material around the symmetry axis has already reached the outer boundary.

%The top right panel shows a snapshot after 2000 yr from the beginning of the simulation. At this point, the shock has completely crossed the initially freely expanding ejecta, heating it up and largely enhancing its expansion rate (but in the incoming jet direction) due to the transfer of energy from the jet to the shocked ejecta. As a consequence, the interaction cross-section increases largely, with the growth of Rayleigh-Taylor instabilities at the shocked jet-ejecta boundary showing large amplitudes.
The top right panel shows a snapshot at $t = 2300$ yr. 
%Now, the backward shock has completely crossed
The backward shock (red/orange transition) continues its propagation through the initially freely expanding ejecta, compressing it and heating it up. It thus serves as a means to transfer kinetic energy from the jet into internal energy through the shocked ejecta. The heating favors the lateral expansion. As a consequence, the shocked jet/ejecta boundary (see the black contour in the lower half panel) greatly increases its cross-section, stops the advance in the upstream direction and develops Rayleigh-Taylor instabilities.

%The bottom left panel shows a snapshot at $t=3000$~yr, with the complete disruption of the bubble and an interaction region now $96$~pc across that almost fills the grid. The quick expansion and instability growth is followed by the disruption of the shocked SN ejecta, and it is dragged and mixed with the shocked jet flow. The instability growth is fast already in 2D, as suggested by previous works studying similar scenarios \citep[e.g.,][]{bosch12,perucho17}.

The bottom left panel shows a snapshot at $t=3000$~yr. At this point, the remnant of the ejecta has been almost stripped by the shocked jet flow and its material dragged downstream and mixed under the action of Rayleigh-Taylor and Kelvin-Helmholtz instabilities. The cross-section of the shocked jet/ejecta boundary now reaches $\sim 60$ pc and the jet bow shock touches the radial boundary of the numerical grid. The instability growth leading to the ejecta remnant disruption is fast even in this axially symmetric 2D case, as suggested by previous works studying similar scenarios, such as in for instance \cite{bosch12}, \cite{perucho17}, and \cite{vieyro19}. In the case of \cite{vieyro19} in particular, in which the interaction of an SN ejecta with a jet was specifically studied, the results qualitatively match those found here. However, the resolution of the S1 2D simulation is $10$ times higher than in the one adopted in \cite{vieyro19} (in which the initial ejecta radius was $\sim R_{\rm max}$, among other more minor differences), allowing for a much more nuanced description of the development of instabilities. This can be seen comparing fig.~A.1 in that work with our Fig.~\ref{2d_timeframe_hr}. Those simulations also addressed a much more unlikely type of event, taking place much closer to the base of a more powerful AGN~jet, so the present results are more relevant for the jetted AGN population as a whole.

The bottom right panel shows the situation at the end of the simulation ($t = 3800$~yr), with the disrupted shocked flow made of mixed jet and ejecta materials being advected down out of the grid but for chunks of shocked ejecta around the axis (probably an artifact of the axisymmetry of the simulation) where the density remains around $10^{-25}-10^{-26}$~g~cm$^{-3}$, that is, two to three orders of magnitude smaller than the initial bubble density, and four or five orders of magnitude denser than the original jet flow. The mixed jet and ejecta material is expected to mass-load the whole jet downstream, outside the grid, forming a tail of turbulent flow somewhat slower than the jet that is likely to homogenize on scales of the jet scale height. %before a conical jet significantly dilute the whole flow. 

%The resulting velocity in the region is $\approx0.5\,c$, well below that of the unshocked jet.

%When most of the shocked ejecta has been pushed downstream and out of the grid, the cross-section of the interaction region is now reduced to that associated to the remains of the SN ejecta on the jet axis, which have still not been pushed by the jet flow. The density of this gas remains around $10^{-25}-10^{-26}$~g~cm$^{-3}$, that is, two to three orders of magnitude more dilute than the initial bubble density. As shown below, the endurance of this structure on the jet axis is probably an artifact of the axisymmetry of the simulation.

At this point, it is interesting to note that the interaction of the jet with the SN ejecta studied in this simulation represents a transient episode in the jet's lifetime since the jet injects amounts of energy and mass equal to those delivered by the SN ejecta in about one and $10^4$ yr, respectively (the latter being similar to the overall interaction timescale).

\subsubsection{3D case}

Figure~\ref{3d_timeframe_rho_hr} shows six snapshots of the density in the $XY$-plane at the middle point of the $Z$-axis ($z=0$). The snapshots show the density distribution on that plane at different times during the simulation, ordered from top left to bottom right. The chosen snapshots correspond to the times shown in Fig.~\ref{2d_timeframe_hr} ($t \simeq 1600, 2300, 3000, 3800$ yr) plus an initial and a final snapshot ($t \simeq 180$ yr and 4700 yr, respectively). During the initial phase (first two panels, $t \leq 1600$ yr) the structure of the ejecta is fairly symmetric (with Rayleigh-Taylor instabilities developing at the shocked jet/ejecta boundary) and resembles the 2D simulation, although the 3D nature of the flow (introduced by the initial perpendicular velocity $v^x_{\rm orb}$ given to the ejecta) prevents the tail of ejecta material around the $Y$-axis (see first panel of Fig.~\ref{2d_timeframe_hr}) to be formed. 

Three-dimensional effects become more apparent as time goes on. Already in the third top panel of Fig.~\ref{3d_timeframe_rho_hr} ($t \simeq 2300$~yr), one sees evidence of turbulent mixing (driven by a combination of Rayleigh-Taylor and Kelvin-Helmholtz instabilities) in the whole region within the jet shock, which has widened up extending over most of the grid and reaching a scale of $\sim 100\,{\rm pc}$ while the shocked ejecta is being quickly disrupted. All this happens significantly faster than in 2D (compare the panels at $t \simeq 2300$ yr in Figs.~\ref{2d_timeframe_hr} and \ref{3d_timeframe_rho_hr}) because of the higher dimensionality and consequent enhanced instability growth. 

%Once the extended shocked region is swept downstream, the shocked region remaining within the grid narrows again, as observed in the bottom center and right panels, but the structure is still fairly disrupted, and there is not a so compact concentration of shocked ejecta as seen on the grid axis of the 2D S1 simulation. In 3D S1, as the remains of shocked ejecta within the grid are still being disrupted by the shock, they generate an interaction area that is also somewhat larger in 3D than in  2D.
As in the 2D case, beyond $3000$ yr (bottom panels of Fig.~\ref{3d_timeframe_rho_hr}) the stripped material from the shocked ejecta starts to be dragged downstream as the shocked region narrows. At $t\simeq 3800$ yr (the last snapshot comparable with the 2D case), the overall structure of the shocked flow in the two simulations is similar. 

%%%%%%%%%%%%%%%%%%%%%%%%%%%%%%%%%%%%%%%%%
%
\begin{figure*}
\centering
\includegraphics[width=0.343\linewidth]{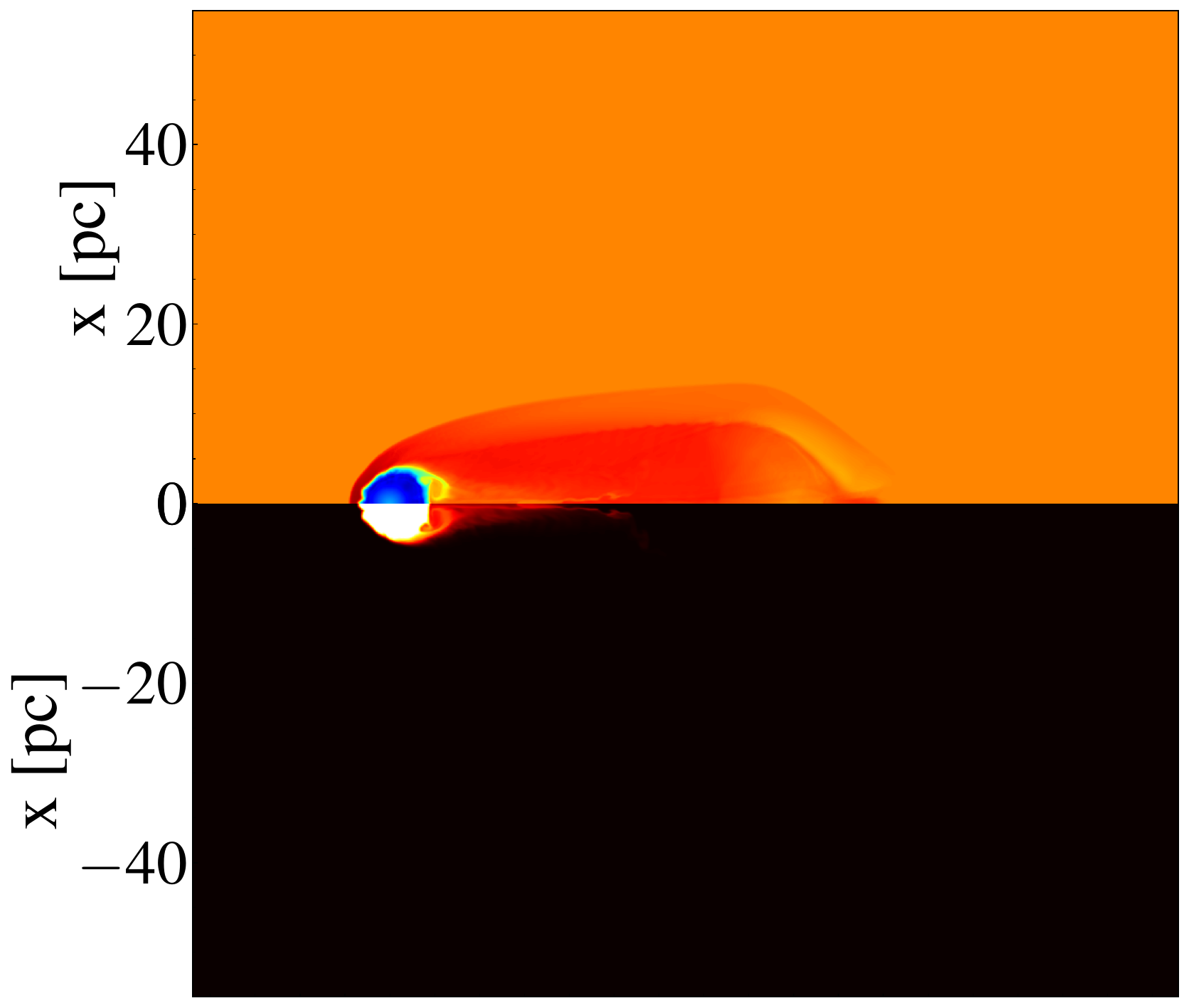}
\includegraphics[width=0.29\linewidth]{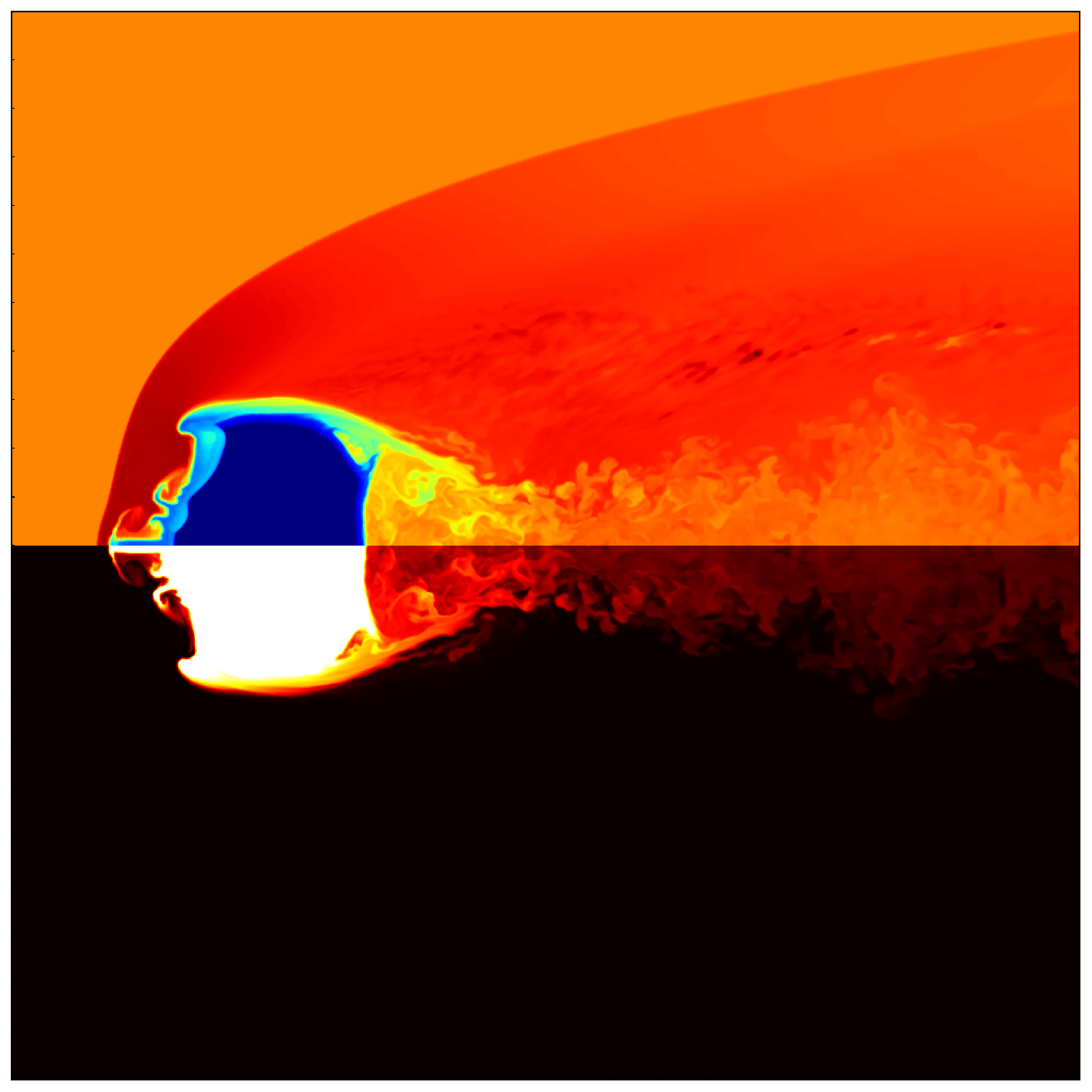}
\includegraphics[width=0.35\linewidth]{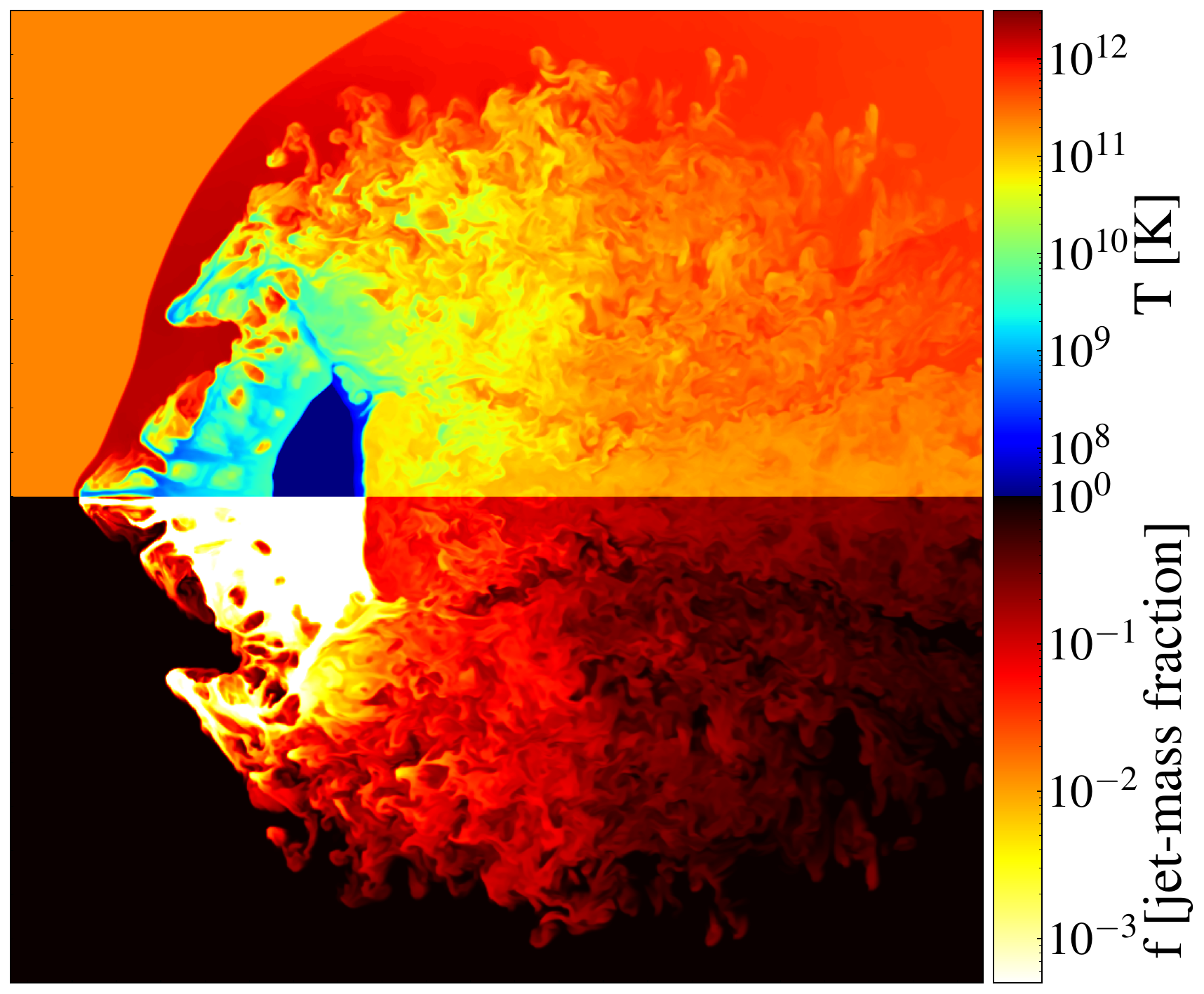}
\includegraphics[width=0.343\linewidth]{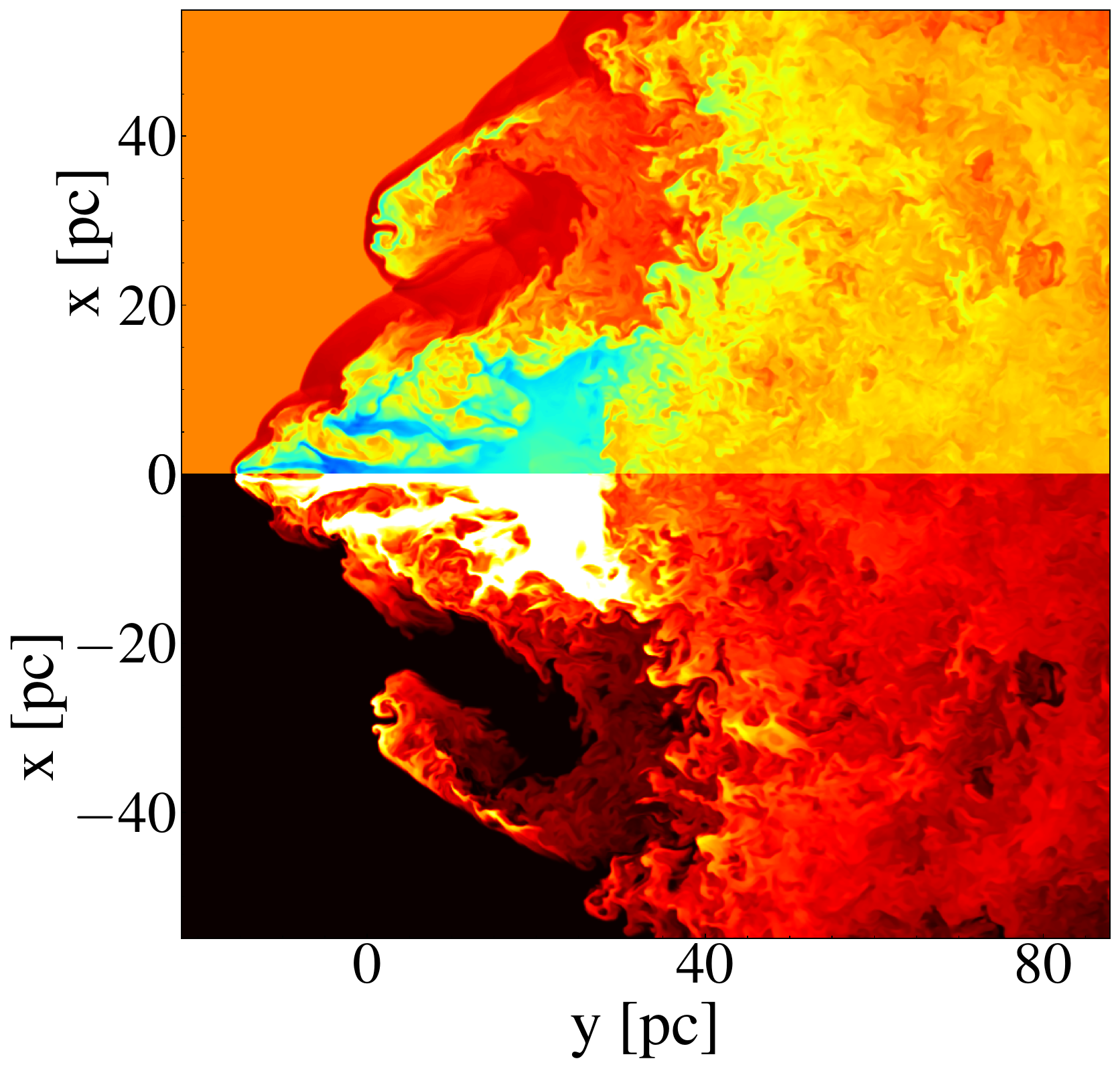}
\includegraphics[width=0.29\linewidth]{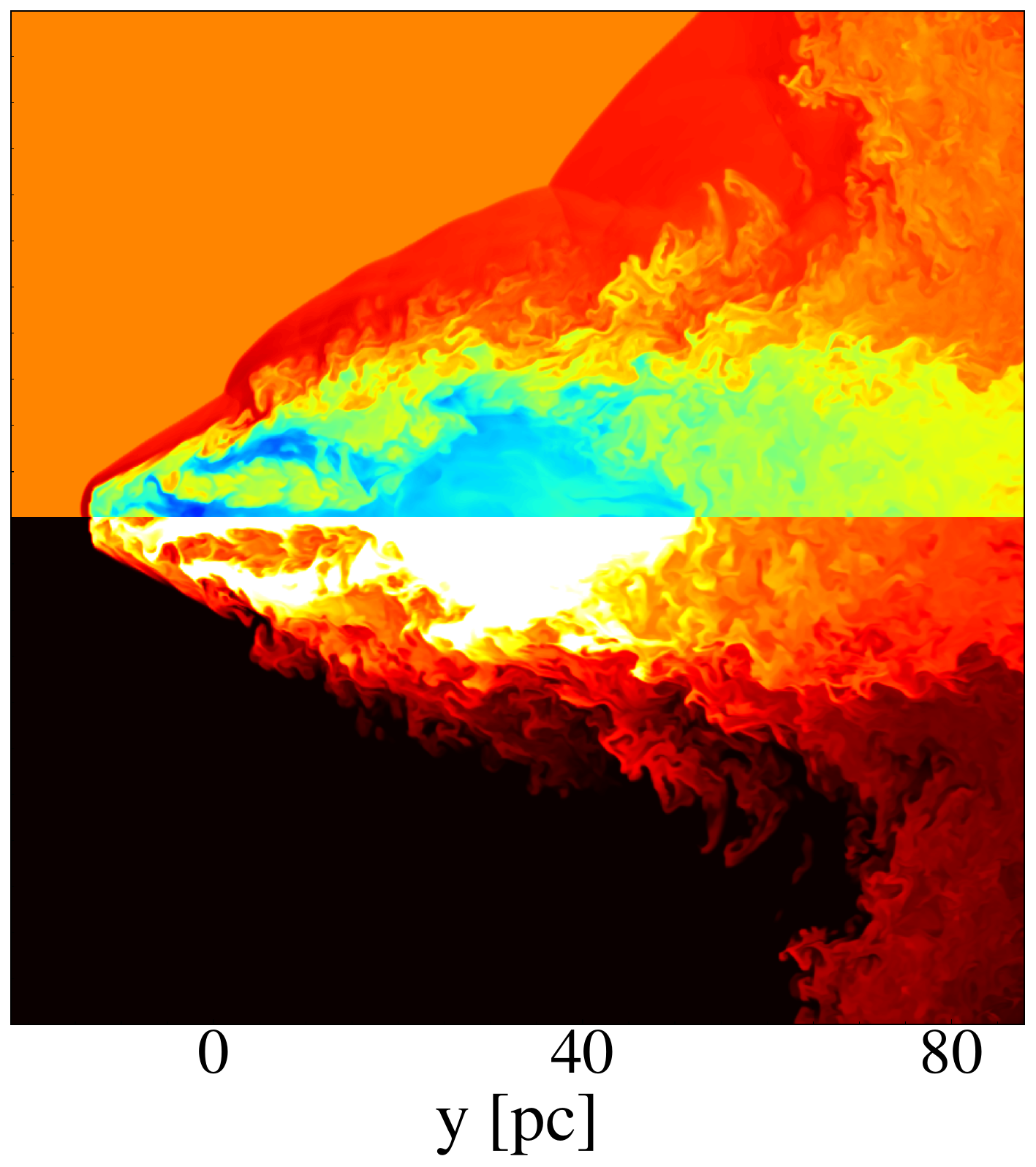}
\includegraphics[width=0.35\linewidth]{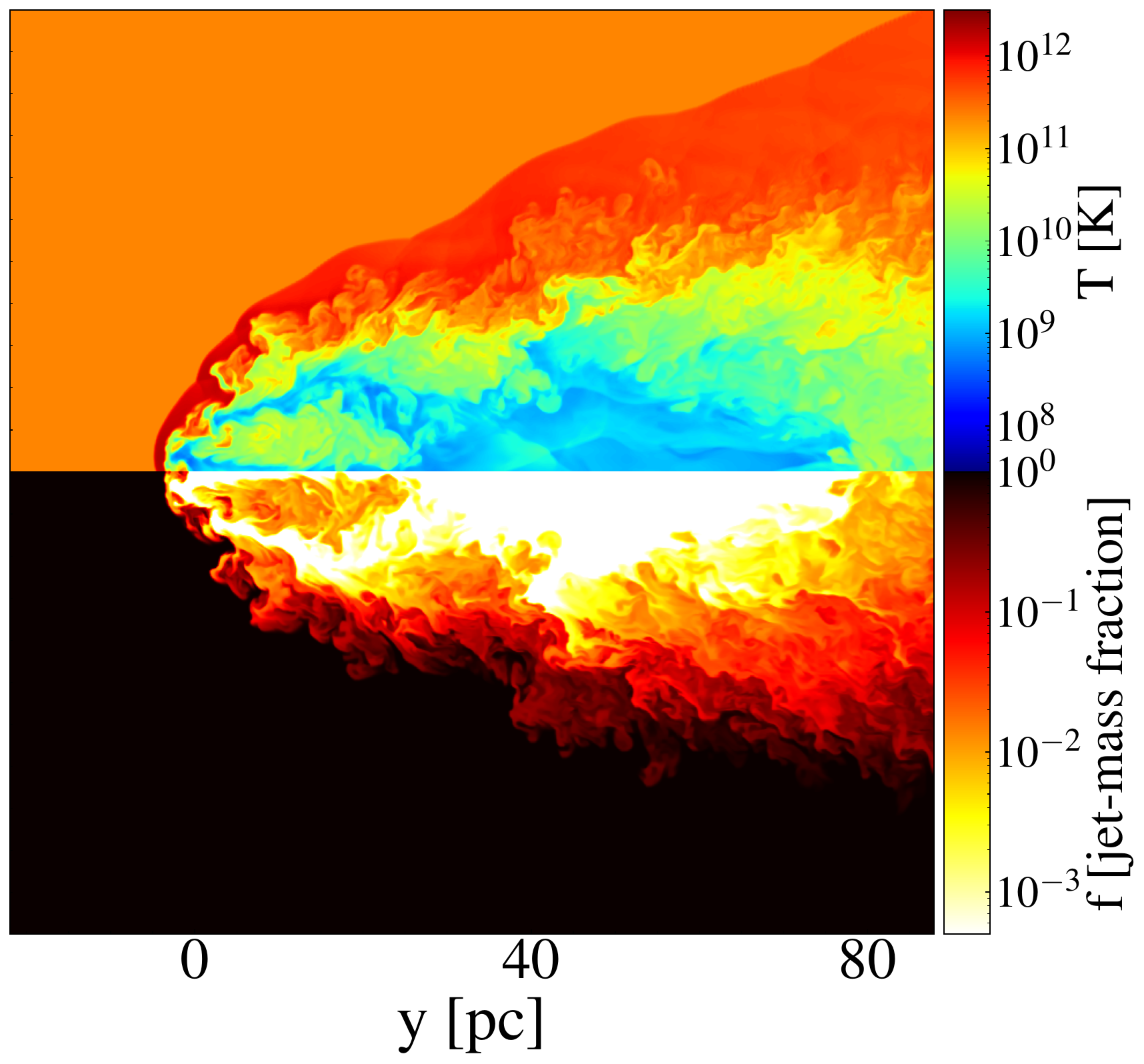}
\caption{Time evolution of the temperature (upper half panels) and the tracer of jet mass-fraction (lower half panels) for S1 in 3D, with 6 2D cuts in the $XY$-plane and at $z=0$; the lower half corresponds to the same region shown in the upper half, but inverted with respect to the axis on $x=0$. The jet is propagating from the left to the right. Top left: $t\approx 180$~yr; top middle: $t\approx1600$~yr; top right: $t\approx2300$~yr; bottom left: $t\approx3000$~yr; bottom middle: $t\approx 3800$~yr; bottom right: $t\approx 4700$~yr.}
\label{3d_timeframe_tem_trac_hr}
\end{figure*}
%
%%%%%%%%%%%%%%%%%%%%%%%%%%%%%%%%%%%%%%%%%%

Figure~\ref{3d_timeframe_tem_trac_hr} shows cuts of the 3D box in the $XY$-plane at $z=0$ for temperature (upper half panels) and tracer (jet mass fraction, lower half panels) for the same times as those in Fig.~\ref{3d_timeframe_rho_hr}. Regarding temperature, the ejecta cools down from the initial $\sim10^{9}$~K to $\leq 10^{8}$~K during the free expansion phase, whereas it reheats dramatically in its shocked phase. The upper panels show the propagation of the initial, jet-driven, shock wave through the cloud and the resulting reheating of the shocked ejecta by two orders of magnitude, reaching temperatures $\sim10^{10}$ K. In contrast, in the much more diluted shocked jet gas the temperature rises up to $\sim10^{12}$~K. The highest temperatures are reached where the jet shock surface is nearly perpendicular to the undisturbed jet flow.

The lower halves of the panels in Fig.~\ref{3d_timeframe_tem_trac_hr} show the jet-mass fraction. The jet material engulfed by the shocked ejecta during the development of the Rayleigh-Taylor instabilities ($t \geq 1600$ yr) favors rapid mixing. At the same time, the development of Kelvin-Helmholtz instabilities at the shocked jet/ejecta interface (already visible at the back of the ejecta beyond $1600$ yr) enhance the mixing rate while the jet bow shock reaches its largest cross-sectional size. Due to the much higher density of the ejecta, the jet-mass fraction remains very low in the inner-most regions of the interaction structure. Nevertheless, the strong mixing observed in the widening two-flow boundary layer (black/red/yellow/white transition) indicates that the jet and the ejecta will mix completely farther downstream. Strong mixing is also visible in the remains of the ejecta that still survive at the end of the simulation. 

%%%%%%%%%%%%%%%%%%%%%%%%%%%%%%%%%%%%%%%%%%%%%%%%%%%%%%%
%
\begin{figure*}
\centering
\includegraphics[width=0.8\linewidth]{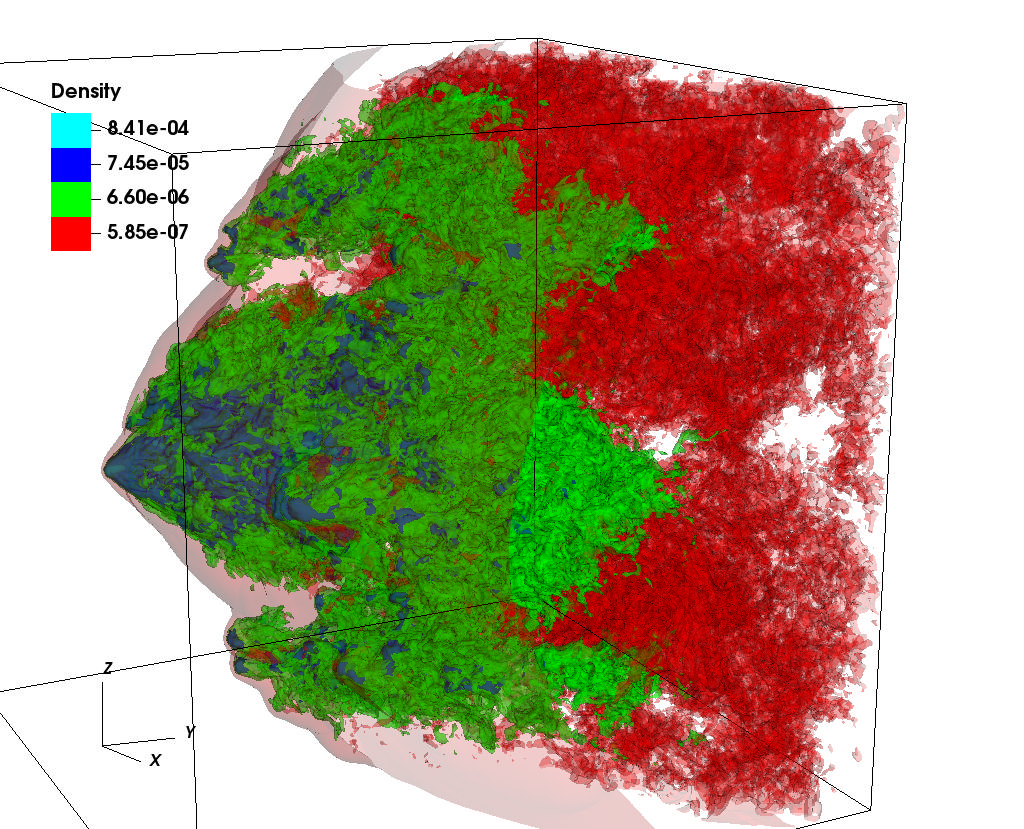}
\caption{Colored surfaces of the rest-mass density 3D distribution, in code units, for the evolved structure of simulation S1 at $t\approx3000$~yr, corresponding to the bottom left panel of Fig.~\ref{3d_timeframe_rho_hr}. The axis of the jet is in the middle of the $XZ$-plane, and the latter propagates along the $Y$-axis (triad visible on the bottom left of the image) and from the left to the right; the surfaces show the shocked ejecta interacting with the jet flow, while the nearly transparent red layer surrounding the shocked ejecta outlines the shocked jet region.}
\label{3d_density_HR}
\end{figure*}
%
%%%%%%%%%%%%%%%%%%%%%%%%%%%%%%%%%%%%%%%%%%%%%%%%%%%%%%

Finally, Fig.~\ref{3d_density_HR} shows the 3D distribution of the shocked structure density in code units, at $t\approx3000$~yr. In the image, the jet flow propagates from left to right, along the $Y$-axis. In red and green we show the surfaces that trace the distribution of the expanded, shocked SN ejecta, whereas the dark blue shows more compact regions, and the light red, partly transparent surface traces the shocked jet layer surrounding the shocked ejecta. The snapshot time corresponds to the maximum shocked structure expansion during the simulation, which coincides with the disruption of the shocked SN ejecta.

\subsection{S2 simulations}
\label{s2}

In the S2 set of simulations, the jet and ejecta properties are the same as in S1 but $R_{\rm SN}$, the initial radius of the supernova ejecta, which is now 2.2~pc. Since the number of cells for the initial radius is still $8$, as in S1, the resolution is effectively twice lower in S2, and its results are slightly different from those of S1. 

\subsubsection{2D axisymmetric case}

\begin{figure*}[]
\centering
\includegraphics[width=0.3952\linewidth]{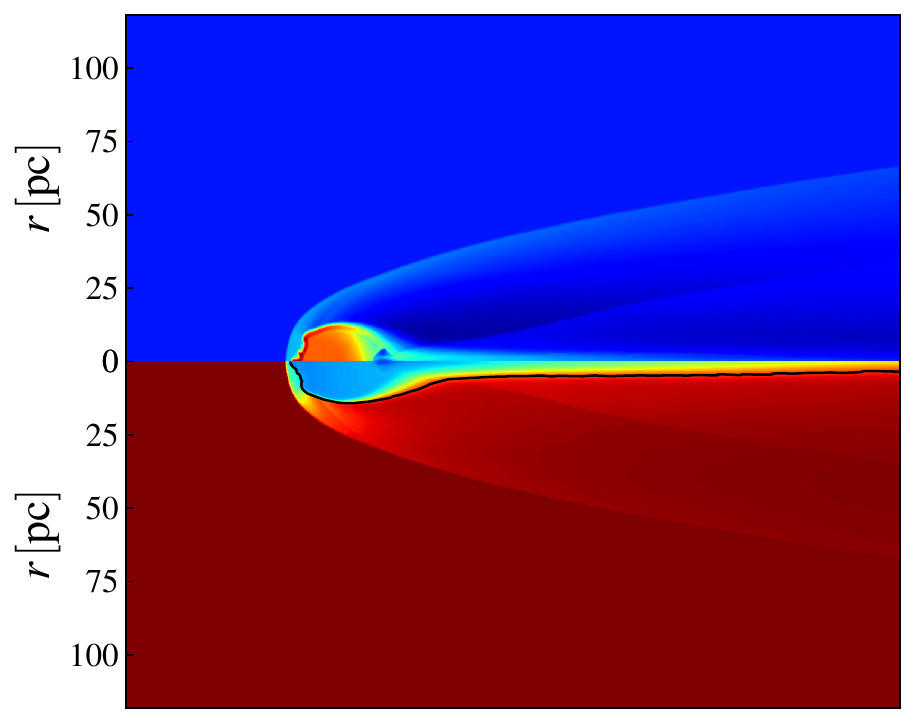} 
\includegraphics[width=0.43264\linewidth]{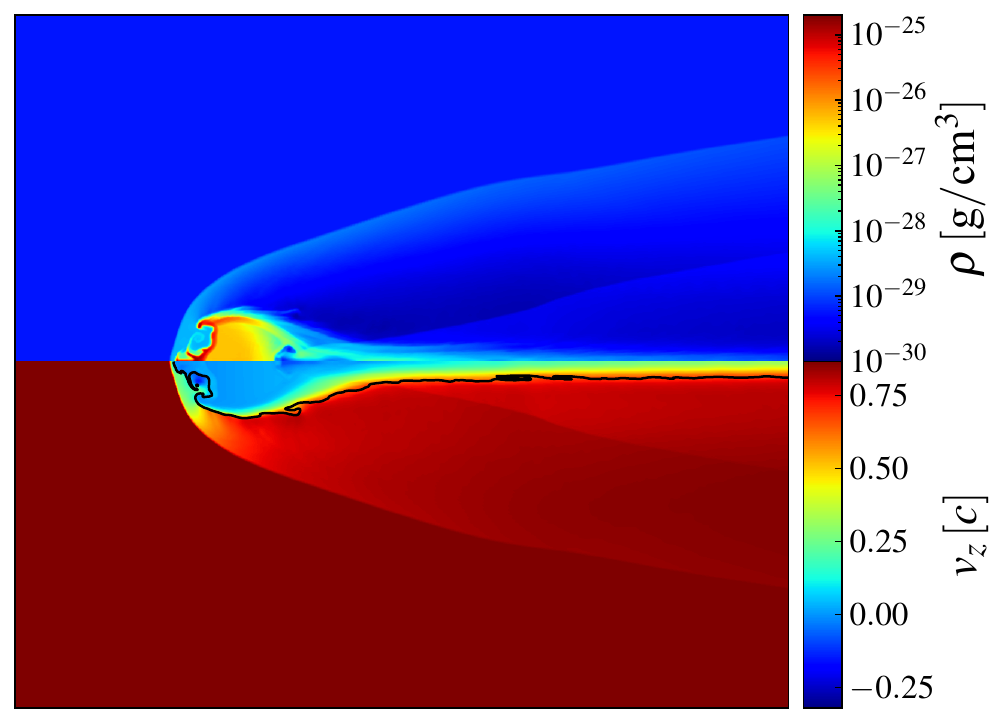}
\includegraphics[width=0.3952\linewidth]{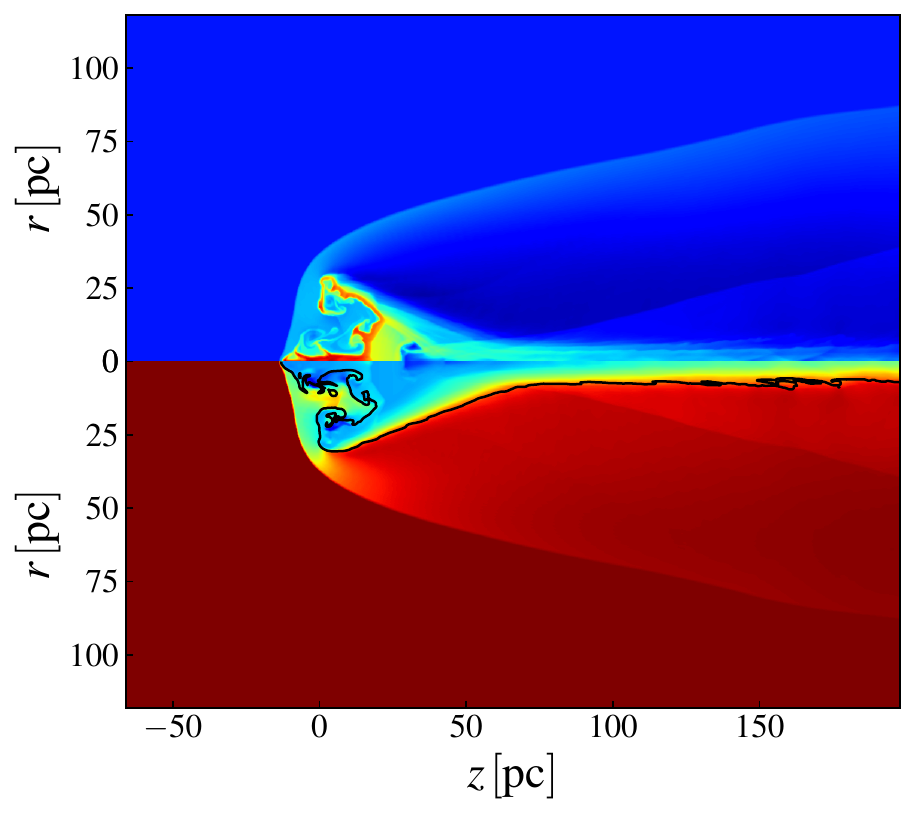} 
\includegraphics[width=0.43264\linewidth]{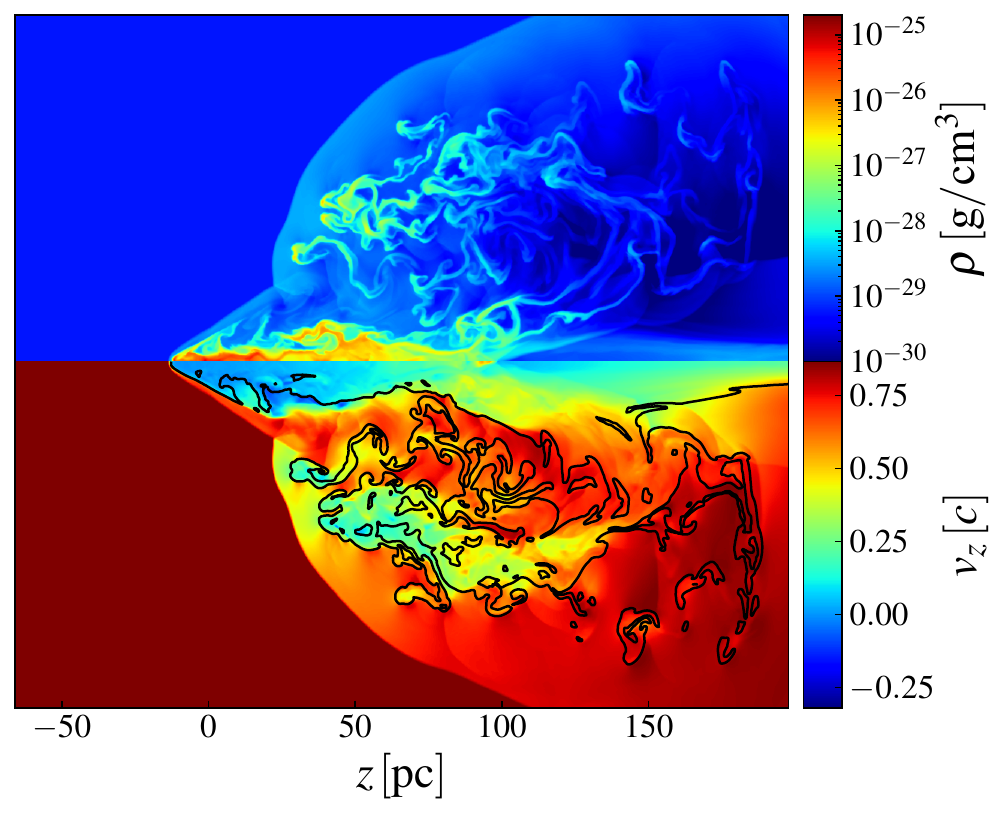}
\caption{Same as Fig.~\ref{2d_timeframe_hr}, but for the 2D axisymmetric S2 simulation. Top left: $t\approx1500$ yr; top right: $t\approx2200$ yr; bottom left: $t\approx2900$ yr; bottom right: $t\approx4000$ yr.}
\label{2d_timeframe_rho_lr}
\end{figure*}

Figure~\ref{2d_timeframe_rho_lr} shows four snapshots of the rest-mass density (upper half panels) and the axial velocity (lower half panels) for the S2 2D simulation at times $t \simeq 1500, 2200, 2900, 4000$ yr. Taking into account that the SN ejecta expands at a speed of $\sim 0.01$~pc/yr, these correspond approximately to those of Fig.~\ref{2d_timeframe_hr}; the SN ejecta is initially more diluted and colder in S2, so the phase of ejecta free expansion captured by the simulation is slightly shorter than in S1 (about 100 yr). In contrast to S1, in which the total disruption of the SN ejecta 
%requires $\sim 2000~\rm{yr}$, in S2 it happens after $\sim 3200~\rm{yr}$,
occurs between $2300$ and $3000$ yr (second and third panels of Fig.~\ref{2d_timeframe_hr}), for S2 this happens between $t= 2900$ and $4000$~yr (third and fourth panels of Fig.~\ref{2d_timeframe_rho_lr}). Most of the mixing takes place during this period, and the maximum lateral expansion of the shocked flow occurs around $\sim 4000~\rm{yr}$ after the start of the simulation, versus $\sim 3000$~yr in S1. Despite this shift caused by the lower resolution of S2, which slows down the development of instabilities, the evolution of the system is qualitatively the same as that observed in S1.

%%%%%%%%%%%%%%%%%%%%%%%%%%%%%%%%%%%%%%%%%%%%%%%%%%%%
%
%
%%%%%%%%%%%%%%%%%%%%%%%%%%%%%%%%%%%%%%%%%%%%%%%%%%%%

\subsubsection{3D case}

%%%%%%%%%%%%%%%%%%%%%%%%%%%%%%%%%%%%%%%%%%%%%%%%%%%%
%
\begin{figure*}[]
\centering
\includegraphics[width=.92\linewidth]{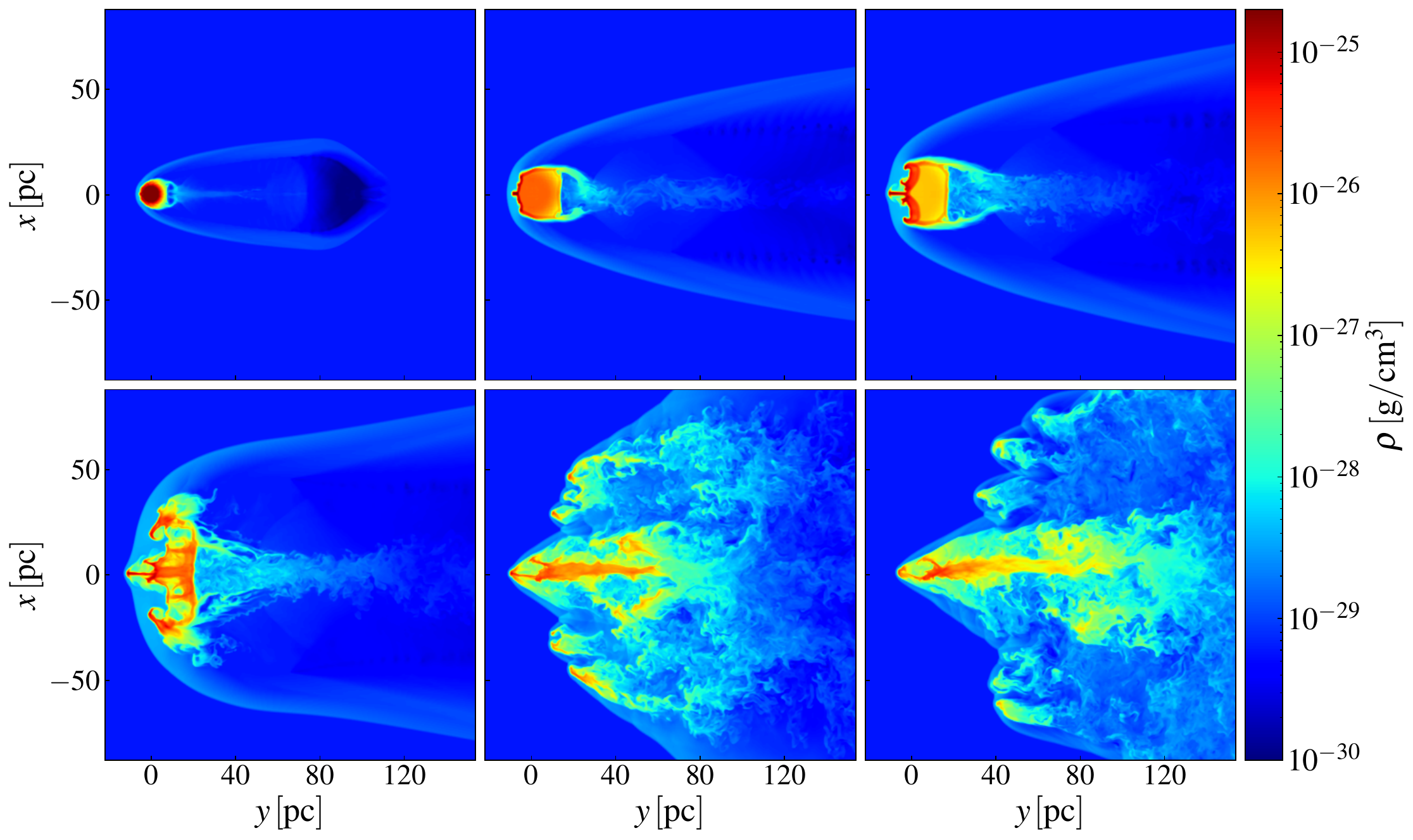}
\caption{Same as Fig.~\ref{3d_timeframe_rho_hr}, for S2 in 3D. Top left: $t\approx350$ yr; top middle: $t\approx1500$ yr; top right: $t\approx2200$ yr; bottom left: $t\approx2900$ yr; bottom middle: $t\approx4000$ yr; bottom right: $t\approx4700$ yr.}
\label{3d_timeframe_rho_lr}
\end{figure*}
%
%%%%%%%%%%%%%%%%%%%%%%%%%%%%%%%%%%%%%%%%%%%%%%%%%%%%%

Figure~\ref{3d_timeframe_rho_lr} shows cuts of the rest-mass density box in the central plane of the $Z$-coordinate ($z=0$) for the 3D S2 simulation. The corresponding times are comparable to those in Fig.~\ref{3d_timeframe_rho_hr}.
As in the case of S1, the break of symmetry introduced by the transversal motion of the SN ejecta becomes evident during the maximum expansion phase of the shocked flow. 
Similarly to the 2D simulations, the total disruption of the SN ejecta is delayed to happen between $2900$ and $4000$ yr (between $t= 2300$ and $3000$ yr in S1). During this phase of disruption, bits of stripped material from the SN ejecta resist the jet ram pressure and propagate up to large distances (see the fifth and sixth panels of Fig.~\ref{3d_timeframe_rho_lr}) before being completely mixed. This also happens in S1, but earlier (see fourth panel in Fig.~\ref{3d_timeframe_rho_hr}). Overall, simulations S1 and S2 follow a qualitatively similar evolution despite the differences introduced by the initial condition, effective resolution and grid size. Interestingly, the larger grid of S2 allows us to follow the evolution of the jet-ejecta interaction for longer times (although the state of the flow in the last frame, $t\simeq 4700$ yr, would be equivalent to a time between $3000$ and $3800$ yr in S1). 

%%%%%%%%%%%%%%%%%%%%%%%%%%%%%%%%%%%%%%%%%%%%%%%%%%%%%%
%
\begin{figure*}[]
\centering
\includegraphics[width=0.343\linewidth]{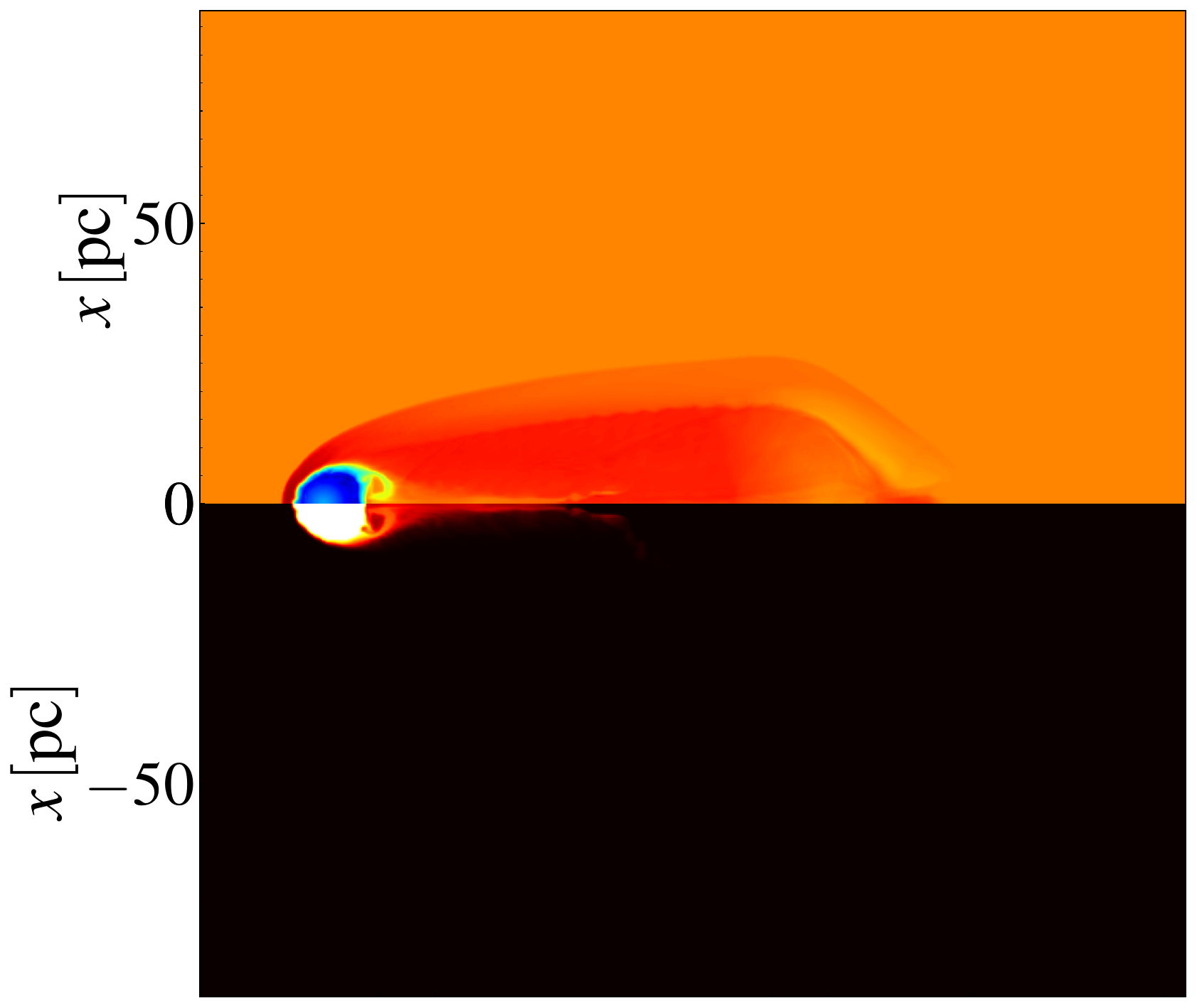}
\includegraphics[width=0.29\linewidth]{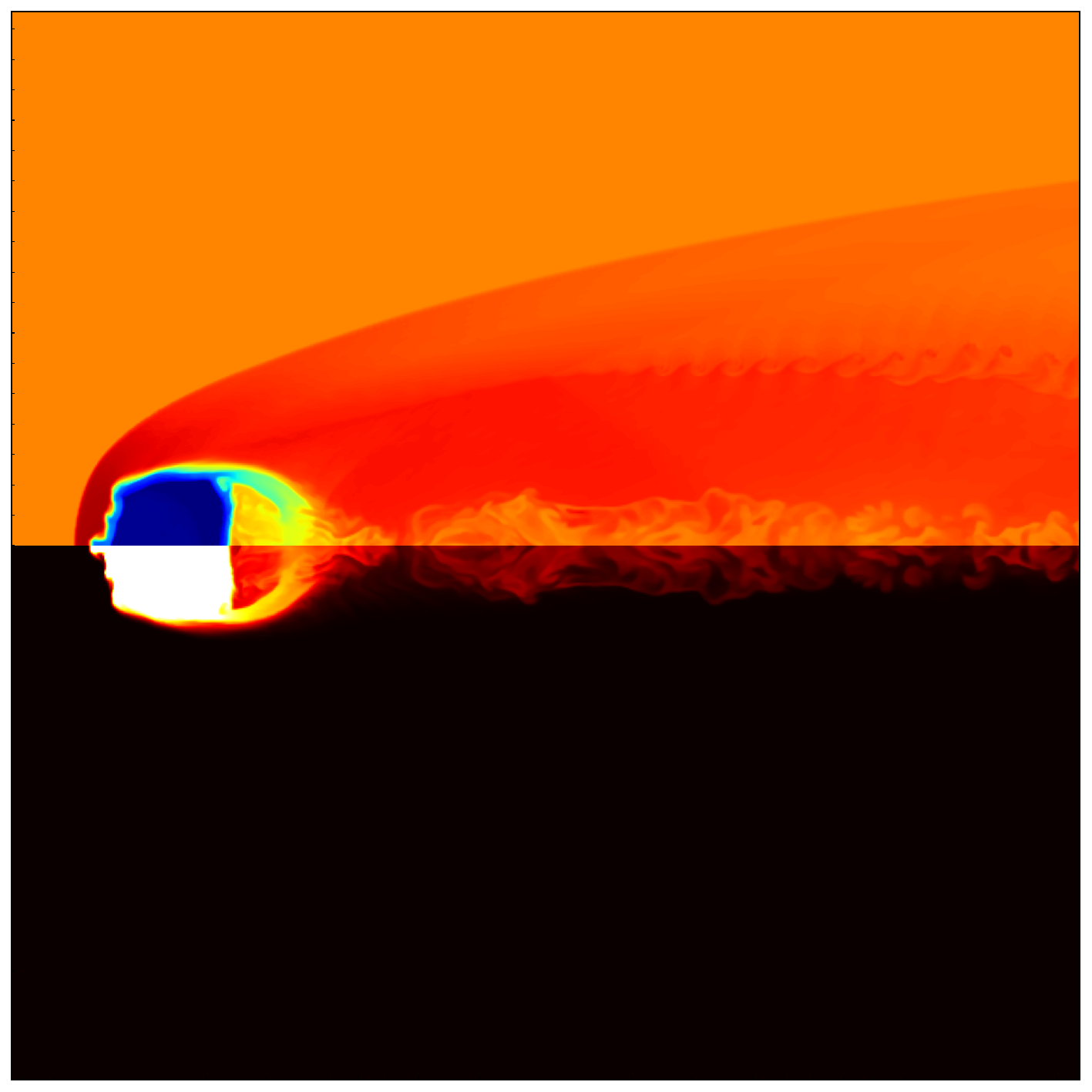}
\includegraphics[width=0.35\linewidth]{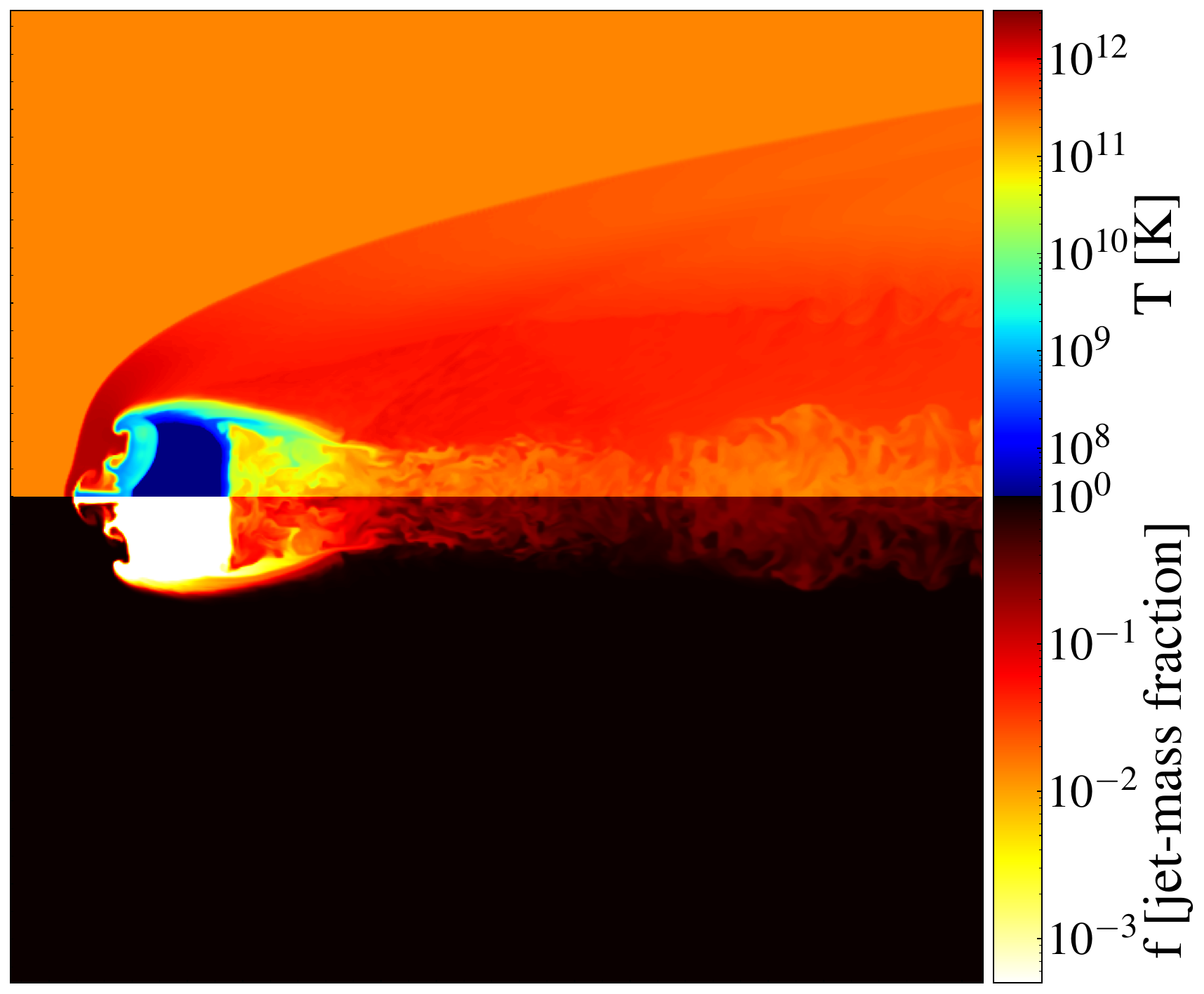}
\includegraphics[width=0.343\linewidth]{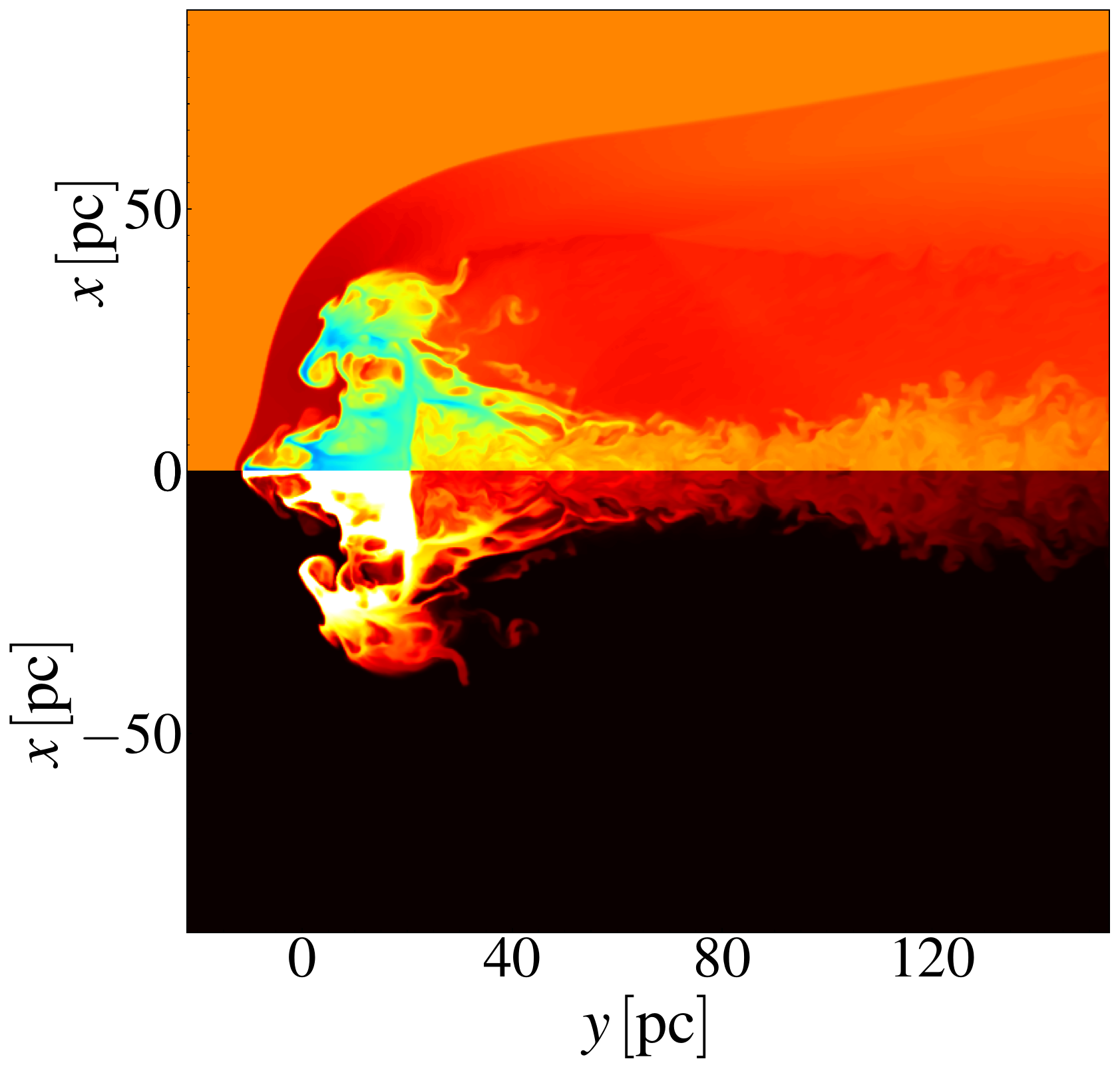}
\includegraphics[width=0.29\linewidth]{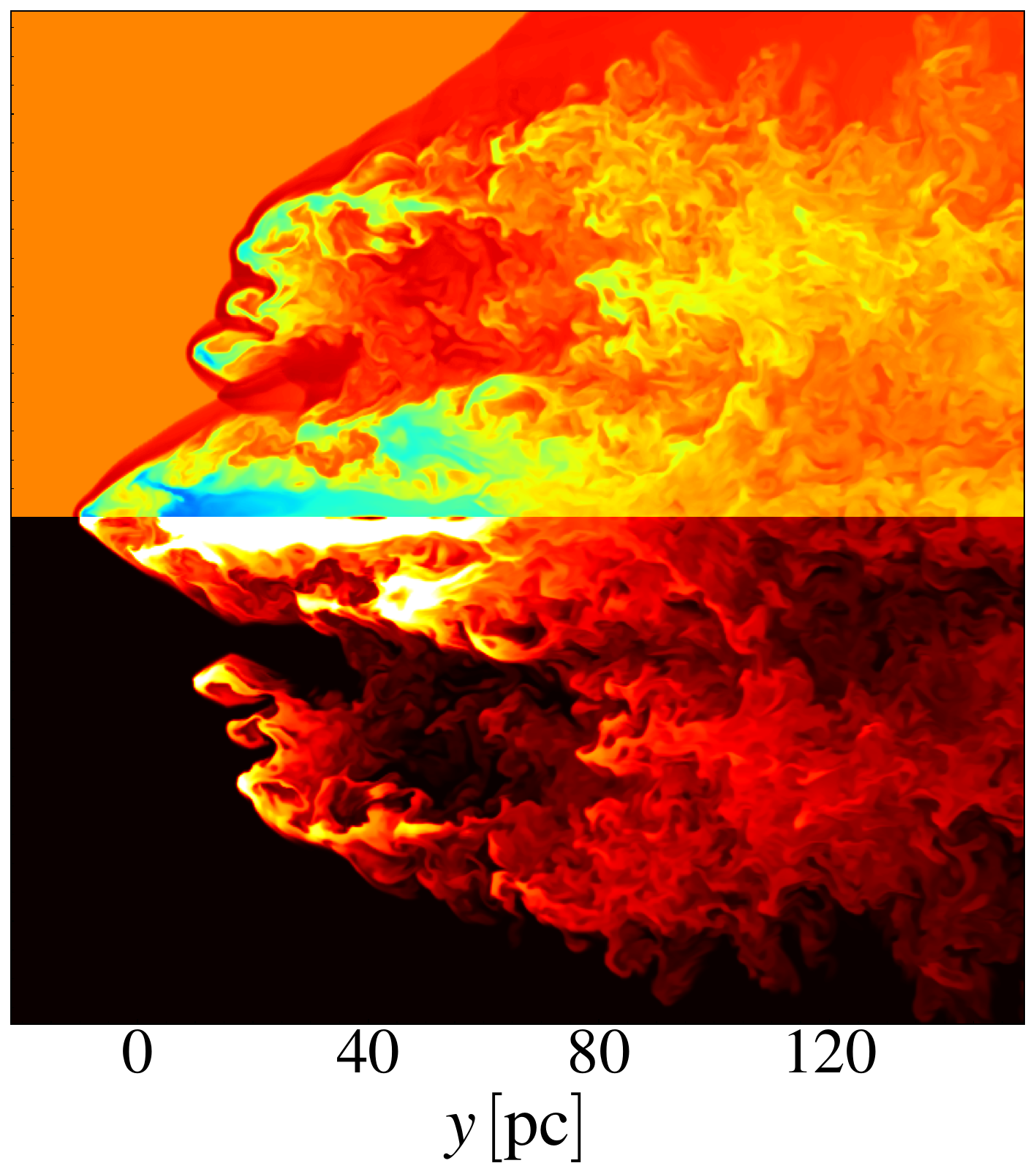}
\includegraphics[width=0.35\linewidth]{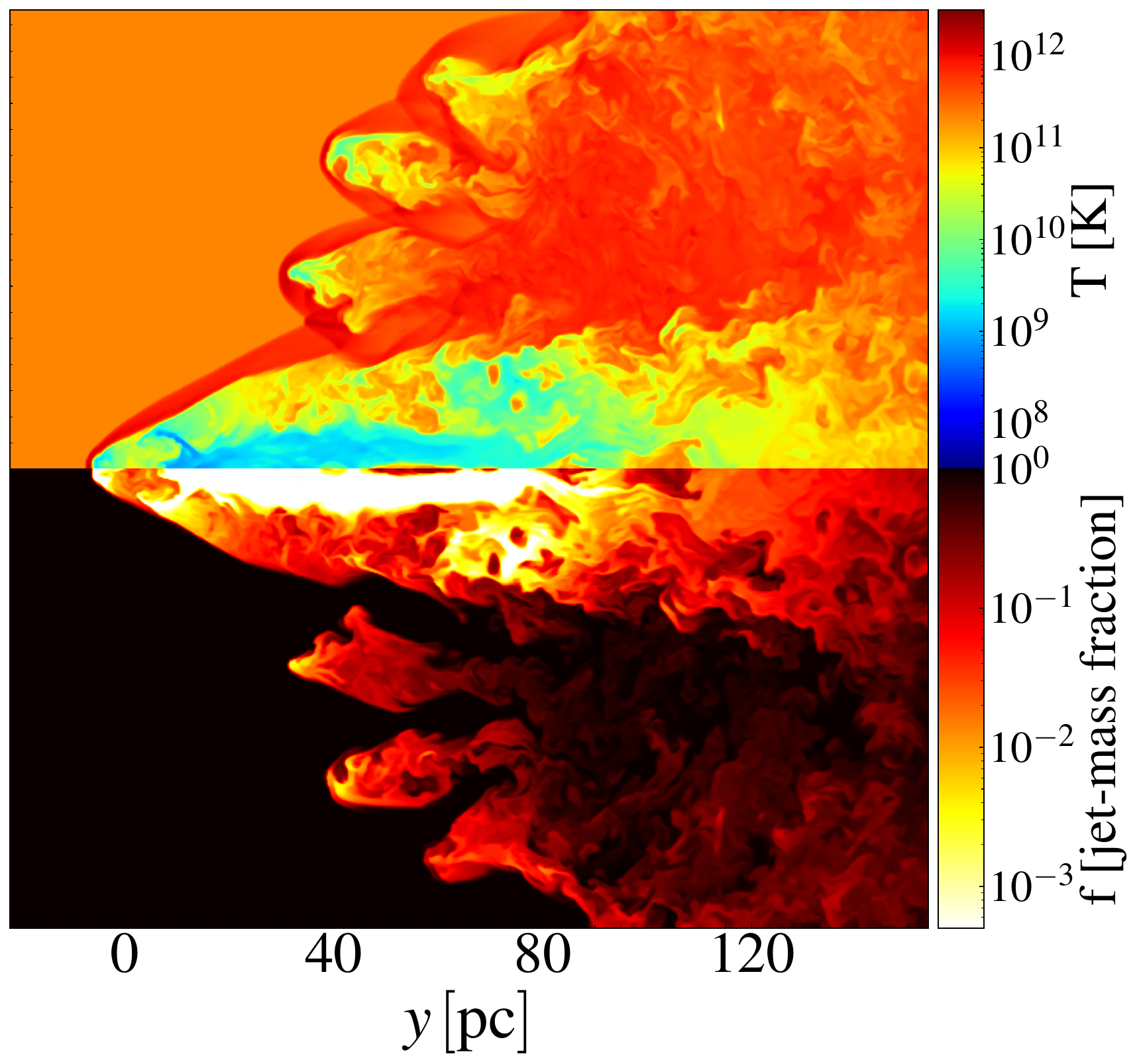}
\caption{Same as Fig.~\ref{3d_timeframe_tem_trac_hr}, for S2 in 3D. Top left: $t\approx350$ yr; top middle: $t\approx1500$ yr; top right: $t\approx2200$ yr; bottom left: $t\approx2900$ yr; bottom middle: $t\approx4000$ yr; bottom right: $t\approx4700$ yr.}
\label{3d_timeframe_tem_trac_lr}
\end{figure*}
%
%%%%%%%%%%%%%%%%%%%%%%%%%%%%%%%%%%%%%%%%%%%%%%%%%%%%

Figure~\ref{3d_timeframe_tem_trac_lr} shows six slices of the temperature (upper half panels) and jet mass fraction (lower half panels) at different simulation times comparable to those in Fig.~\ref{3d_timeframe_tem_trac_hr}. 
%The values of temperature achieved in the post-shock material are similar to those in S1.
%, peaking at $\sim10^{12}$~K for the shocked jet gas, and at $5\times10^{9}-10^{10}$~K for the SN remnant.
The distribution and values of temperature in the post-shock material along the simulation are similar to those in S1. The distribution of the tracer at long times ($t \geq 4000$, comparable with times $t \geq 3000$ in S1) allows us to conclude that although the smaller resolution in S2 delays the development of instabilities, it is more effective in stripping and mixing the ejecta material once they develop to non-linear amplitudes. 
%The figure panels show turbulent mixing and advection of the ejecta remains along the jet, and the interaction structure reaches a size $\sim 100$~pc, as in S1. In S2 the larger grid allows us to follow the jet-ejecta interaction up to slightly larger regions, but aside from the somewhat slower evolution of S2, the differences between the two simulations are not strong.

%%%%%%%%%%%%%%%%%%%%%%%%%%%%%%%%%%%%%%%%%%%%%%%%%%%%%%
%
\begin{figure*}
\centering
\includegraphics[width=0.75\linewidth]{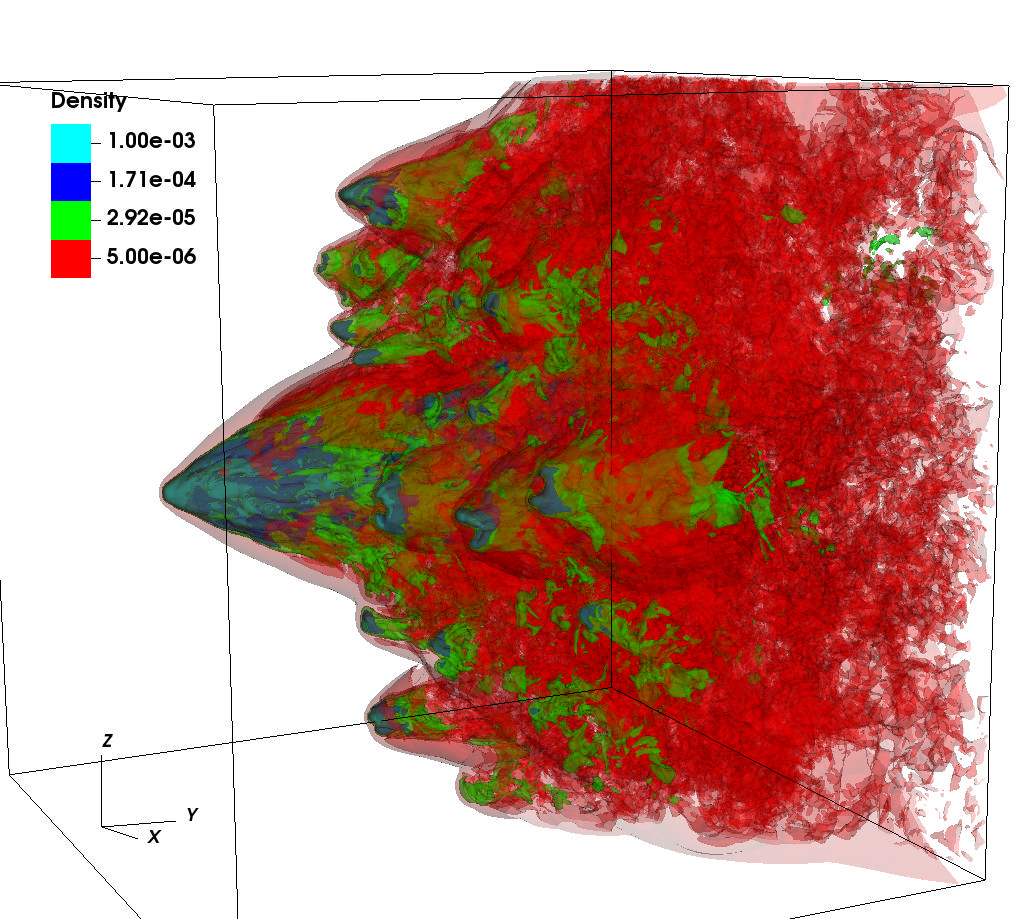}
\caption{Same as Fig.~\ref{3d_density_HR}, for S2 in 3D at $t=4700$ yr.}
\label{3d_density_LR}
\end{figure*}
%
%%%%%%%%%%%%%%%%%%%%%%%%%%%%%%%%%%%%%%%%%%%%%%%%%%%%%%%

Finally, Fig.~\ref{3d_density_LR} shows the equivalent of Fig.~\ref{3d_density_HR} for S2, after $4700$~yr from the start of the simulation. Again, the image shows a qualitatively similar picture to that resulting from S1, with dense knots triggering small-scale interactions as the shocked ejecta expands and gets disrupted. Beyond this region, efficient mixing with the shocked jet gas is also observed, involving scales that become of the order of the jet radius. 

\subsection{Maximum upstream expansion}

\begin{figure*}
\centering
\includegraphics[width=.47\linewidth]{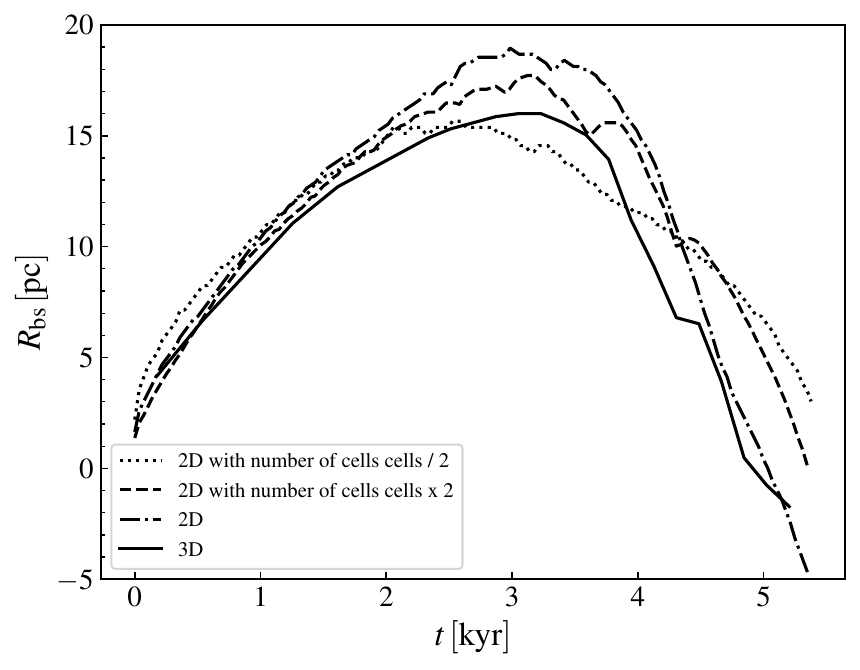}
\includegraphics[width=.47\linewidth]{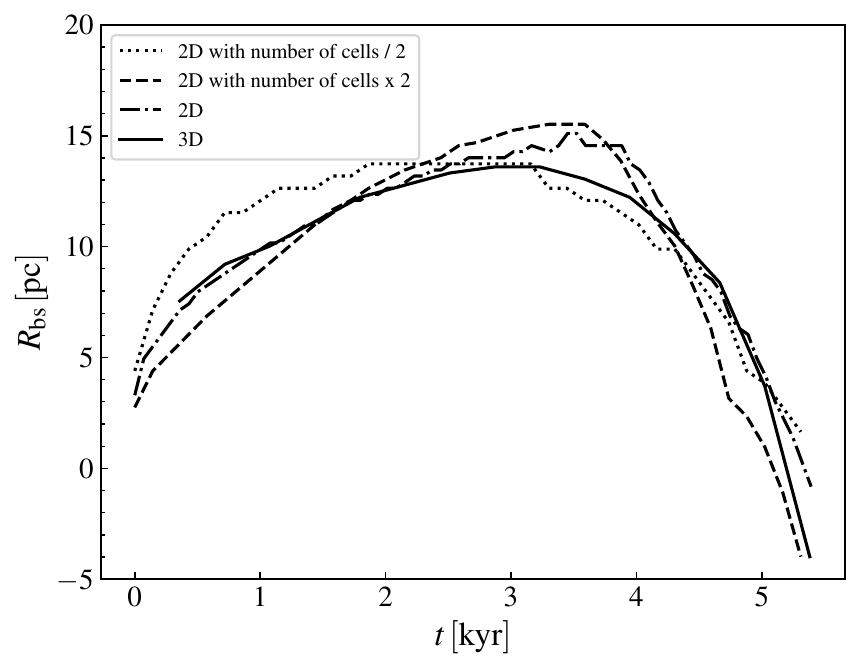}
\caption{Time evolution of the jet bow shock location in S1 (left panel) and S2 (right panel) in 2D and 3D. For the 2D simulations, the calculation is done for 3 different resolutions: same resolution as the 3D simulation, resolution / 2 (i.e., number of cells /2) and resolution $\times 2$ (i.e., number of cells $\times  2$).}
\label{eq_point_s1}
\end{figure*}

In Sect.~\ref{simulations_ref}, we gave an analytical estimate of the position of the equilibrium point between the ejecta and jet ram pressures, as measured from the SN original location, $R_{\rm max}\simeq 11$~pc. Figure~\ref{eq_point_s1} shows the evolution of the position of the jet bow shock, $R_{\rm bs}$, with time along the symmetry axis of the initial ejecta, for the different simulations: S1 (left), S2 (right), 3D, and 2D for different resolutions (half and double of the reference value). The shock position is defined as the first jump in pressure met along the axis. 

The curves in the figure show a fast initial rise of $R_{\rm bs}$, which corresponds to the ejecta free expansion phase for all the studied cases. Around $t\sim 1000$~yr, the jet bow shock location has reached values close to 10 pc. Its displacement slows down, but it keeps advancing upstream up to values well beyond the analytical estimate: in S2, $R_{\rm bs}\approx 13$~pc in 3D and $\approx 15$~pc in 2D; in S1, $R_{\rm bs}\approx 16$~pc in 3D and $\approx 18$~pc in 2D 
(the 2D simulations present an acceptable agreement for the different resolutions).
%(the 2D simulations present an acceptable degree of convergence when increasing the numerical resolution). 
Eventually, the shock starts to recede as the shocked ejecta is pushed downstream of the jet. 

In all cases, the maximum of $R_{\rm bs}$ is achieved $3000-4000$~yr after the beginning of the simulations, much later than the analytical estimate of $\sim 1000$~yr. This long delay, together with the large values achieved by $R_{\rm bs}$ can be understood, beyond the intrinsic complexity of the real jet-ejecta interaction, by considering the following two facts. First, the analytical estimate for the time needed to reach the maximum expansion assumes a constant expansion speed (equal to the initial one), without taking into account deceleration down to zero when equilibrium is reached. And second, the estimate of the position of the jet bow shock location is purely one-dimensional and ignores the fact that since the ejecta has a finite radius, and the jet flow is deflected laterally at the bow shock, there is a drop in the effective ram pressure exerted on the ejecta material. 

The density in the shocked ejecta decreases with time (and hence its inertia) and the jet starts to push the ejecta remnant downstream in an accelerated manner. At $t\simeq 5000$~yr, the shocked jet-ejecta structure crosses the initial center of the SN explosion (bottom right panels of Figs.~\ref{3d_timeframe_rho_hr} and \ref{3d_timeframe_rho_lr}), which is also indicated in Fig.~\ref{eq_point_s1} by the negative values of the position of the jet bow shock. 

%%%%%%%%%%%%%%%%%%%%%%%%%%%%%%%%%%%%%%%%%%%%%%%%%%%%
%
%
%%%%%%%%%%%%%%%%%%%%%%%%%%%%%%%%%%%%%%%%%%%%%%%%%%%%%
%%%%%%%%%%%%%%%%%%%%%%%%%%%%%%%%%%%%%%%%%%%%%%%%%%%%%
%
\begin{figure*}
\centering
\includegraphics[width=.49\linewidth]{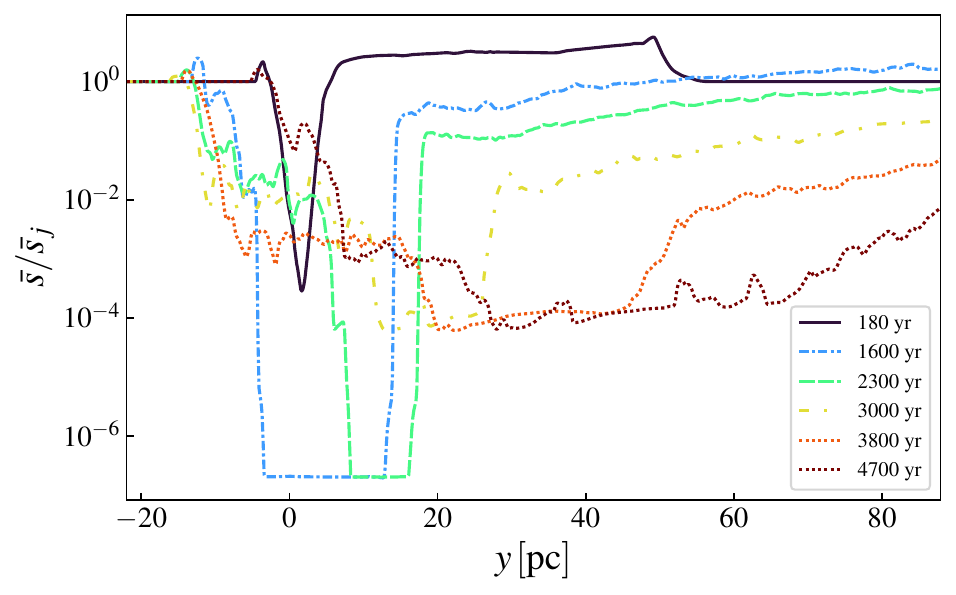}
\includegraphics[width=.49\linewidth]{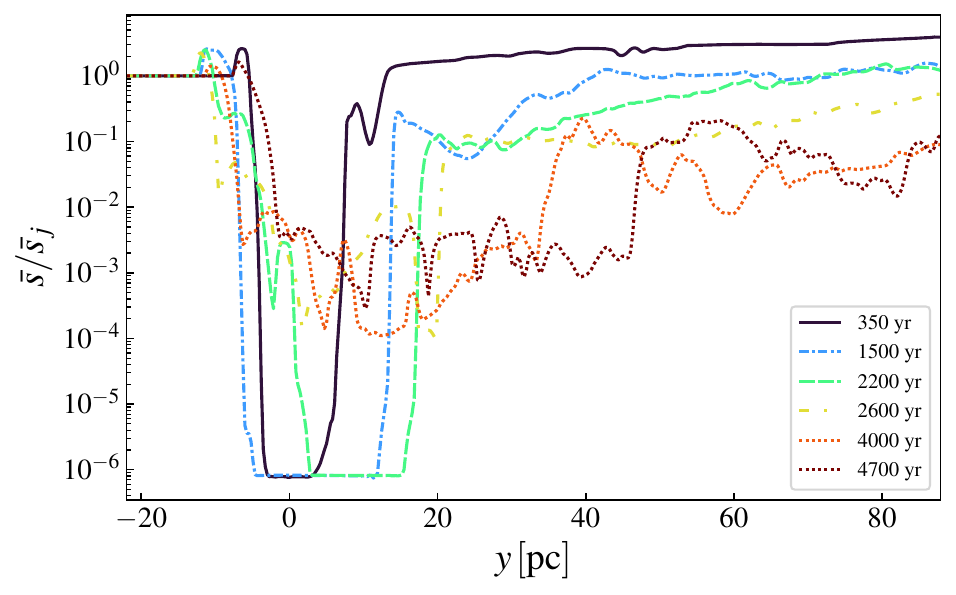}
\caption{Specific entropy (normalized to the jet entropy) averaged over the domain $[-2.5,2.5] \, R_{\rm SN} \times [-2.5,2.5] \,R_{\rm SN}$ of the $XZ$-plane around the jet axis for the 3D simulations; 6 time steps from S1 (left panel) and S2 (right panel) are shown to accurately follow the dynamical evolution of the ejecta.}
\label{integrated_entropy}
\end{figure*}
%
%%%%%%%%%%%%%%%%%%%%%%%%%%%%%%%%%%%%%%%%%%%%%%%%%%%%%%

%%%%%%%%%%%%%%%%%%%%%%%%%%%%%%%%%%%%%%%%%%%%%%%%%%%%%%
%
\begin{figure*}
\centering
\includegraphics[width=.49\linewidth]{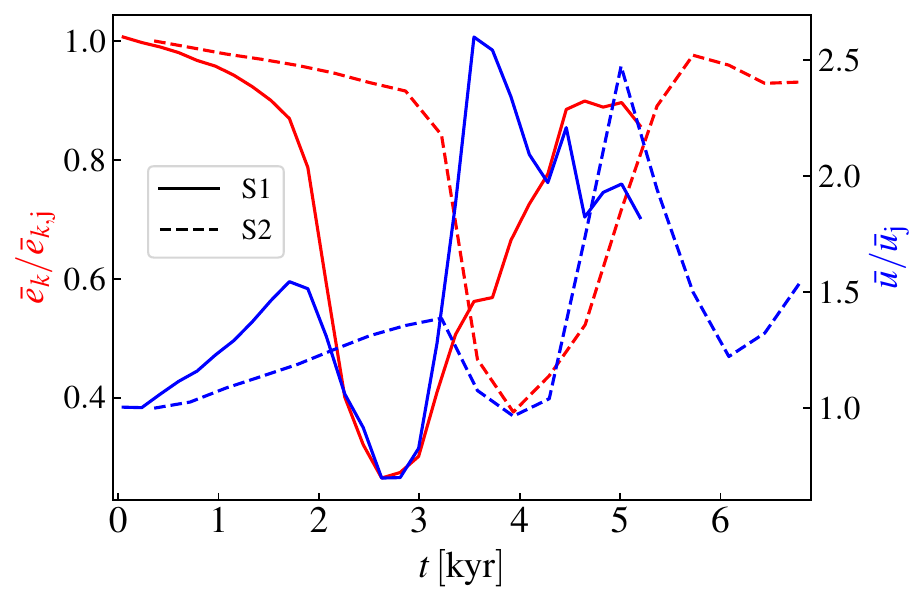}
\includegraphics[width=.49\linewidth]{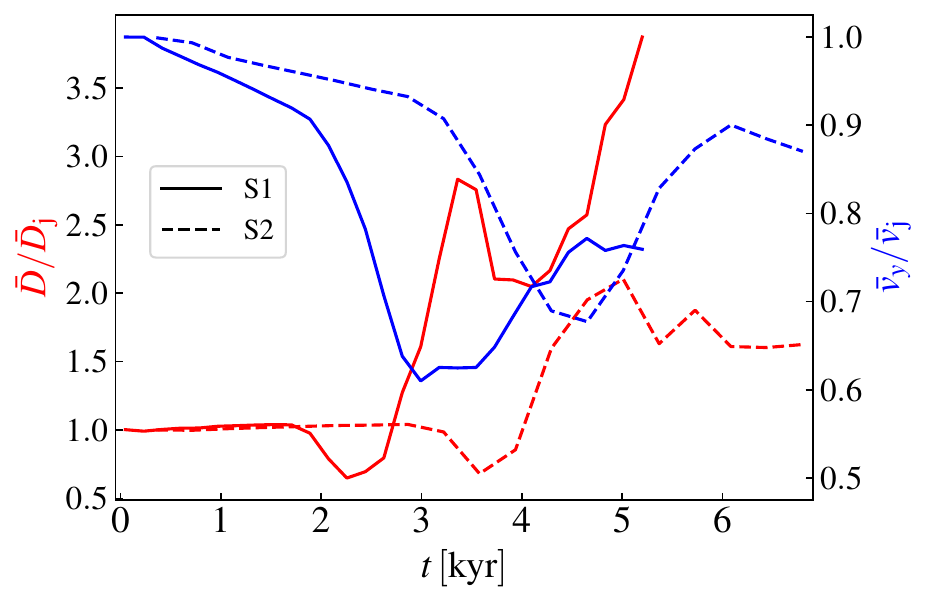}
\caption{Time evolution in the 3D case of the rest-mass density in the lab frame $D$, kinetic $e_{\rm k}$ and internal $u$ energy density (normalized to the jet values), and velocity $v_y$ in $v_{\rm j}$-units, averaged over the $XZ$-plane at the outflow boundary of the $Y$-axis, for S1 (solid lines) and for S2 (dashed lines).}
\label{time_evo_u_uk_d_vy}
\end{figure*}
%
%%%%%%%%%%%%%%%%%%%%%%%%%%%%%%%%%%%%%%%%%%%%%%%%%%%%%%%

\subsection{Dynamics}

Entropy is a good indicator of the changes suffered by the interacting flows as they evolve.
%, so we explored the evolution of entropy in the 3D simulations. 
Ignoring logarithms and constants, we define the (specific) entropy as $s = T\rho^{-2/3}$ \citep[see, e.g.,][]{lloyd00}. 
Figure~\ref{integrated_entropy} shows the average of specific entropy for the 3D simulations S1 and S2 over the domain $\Sigma = [-2.5,2.5] \, R_{\rm SN} \times [-2.5,2.5] \,R_{\rm SN}$ of the $XZ$-plane along the $Y$-axis and normalized to the jet averaged entropy, for different evolution times. Restricting the averaging to $\Sigma$ allows us to focus on the evolution of the entropy within the central region of the SN ejecta.
%and applying it along the jet axis for different evolution times yield the results shown in Fig.~\ref{integrated_entropy}. The value is averaged over the $XZ$-plane taking ([$-2.5,2.5$] $R_{\rm SN}$ $\times$ [$-2.5,2.5$] $R_{\rm SN}$) around the axis on $x,z=0$ in order to limit the calculation to the interaction region. 
%It is computed, for each time step, as:
%\begin{align}
%    \bar{X}=\frac{\int X dS}{\int dS},
%\end{align}
%where $dS=dxdz$.
%The left panel corresponds to S1, whereas the right panel corresponds to S2. 
%In all the cases, the ejecta is seen as a drop in entropy, but entropy rises with time because of the ejecta being shocked and the development of turbulence. 
The specific entropy of the ejecta material, initially small compared with that of the jet, rises with time as the ejecta is swept by the backward shock and turbulence develops. At the same time, the ejecta remnant moves downstream dragged by the jet flow.

We also studied the evolution of the total thermal energy density, $u = \rho\,\,\varepsilon\, c^2\,(\Gamma\,W^2+1-\Gamma)$, the kinetic energy density, $e_{\rm k}=\rho\,c^2\,W\,(W-1)$, the density $D=\rho\,W$, in the lab frame, and the axial velocity, $v_y$, at the right boundary of the grid through which the shocked flow leaves the computational domain. We show the results in Fig.~\ref{time_evo_u_uk_d_vy}. The values of $e_{\rm k}$, $u$, $D$ and $v_y$ are normalized to the background jet ones. After an initial drop caused by a rarefaction wave formed downstream of the ejecta, both quantities increase at the right boundary as the jet-ejecta interaction evolves and the shocked flow is advected by the jet out of the grid. The increase in $e_{\rm k}$ is associated to the ejecta mass incorporated to the jet, whereas the increase in $u$ is driven by shock heating (see Figs.~\ref{3d_timeframe_tem_trac_hr} and \ref{3d_timeframe_tem_trac_lr}).
The right panel of the Figure shows how the right-boundary averaged $D$ initially rises while $v_y$ drops. However, $v_y$ grows again due to acceleration once the ejecta has been disrupted and pushed by the jet after $\sim 3000$ and $4500$~yr in S1 and S2, respectively. The delay observed in S2 with respect to S1 again indicates that a higher resolution slightly increases the pace of the dynamical processes. The value of $v_y$ becomes mildly relativistic at the end of the simulations ($\sim 0.65-0.75\,c$).

\section{Discussion and conclusions}\label{disc}

Our simulations show that a SN explosion can strongly mass-load a relativistic jet on a timescale $\sim 10^4$~yr, consistent with the analytical estimate given in \cite{vieyro19} in the present scenario. This timescale is of the order of the interaction height in the jet over $c$, gets smaller for more powerful jets (less common), and longer for weaker jets (more common). As noted in \cite{bosch23}, the fraction of jetted active galactic nuclei hosting an interaction with a SN should be of the order of the interaction duration times the SN rate within the jet.

There are several effects of the jet-ejecta interaction that are worth mentioning: producing temporary jet deceleration and generation of transient inhomogeneities in velocity along the jet; enriching the outflow with metals that are carried on to the host cluster of the active galaxy \citep[e.g.,][]{2009ApJ...707L..69K,2010MNRAS.405...91S,2020ApJ...904....8C} or can be accelerated to ultra-high energies \citep{bosch23}; and triggering potentially detectable non-thermal emission as a consequence of the conversion of jet kinetic energy into (non-thermal) internal energy at shocks and turbulent regions \citep{vieyro19}. 
%\subsection{Mass-loading and mixing}

Regarding jet mass-load, the mean rate during the simulation time is of a few times $10^{-4}\,M_{\odot}\,$yr$^{-1}$, taking into account the simulation timescale and the initial SN ejecta mass, which can be compared to the mass injection rate of the jet considered, $\sim 6 \times 10^{-4}\,M_{\odot}\,$yr$^{-1}$. Although we model it as a simple electron-proton gas, the ejecta is largely made of heavy nuclei, which will mix with the jet flow as it propagates out of the galaxy. These situations can thus add non-negligible amounts of heavy metals downstream of a SN explosion to those brought by the jet due to regular mass load (e.g., entrained galactic gas, stellar winds, and accretion disk matter); the SNe may even dominate the content of heavy elements in the jet termination regions. 

The simulated mass-loaded jet bulk flow features a transitory deceleration, which, for the typical parameters adopted here, can affect the whole jet cross-section. In this context, clumps of heavier material from the disrupted ejecta would be advected along the jet for a distance, with upstream, faster jet flow catching up these slower regions. The affected jet region would eventually become diluted in the jet flow while propagating towards the jet termination region –hotspot– in Fanaroff–Riley type-II (FRII) radio galaxies, or into the surrounding plume filled by jet-ambient mixed flow in the case of FRIs.

%\subsection{Impact of the spatial scale}

The jet-ejecta interaction was simulated in 2D and 3D, and for different spatial resolutions. The simulations follow a qualitatively similar evolution, which suggests that in the context of (relativistic) hydrodynamical simulations, we are approaching a realistic spatial resolution level. However, although the lower resolution of S2 allowed us to extend the study to larger spatial and temporal scales, the speed of the shocked ejecta disruption process is slightly faster when the resolution is increased, likely related to the lower numerical viscosity in this case (S1). The same result is obtained when moving from 2D to 3D simulations, for which in addition we find that tiny perturbations of the symmetry through $v^x_{\rm orb}$ lead to a stronger impact of instabilities in the long-term evolution of the shocked ejecta, as seen when comparing for instance Fig.~\ref{3d_timeframe_rho_hr}, with Fig.~\ref{3d_timeframe_rho_hr_wm}. The differences introduced by the orbital motion of the ejecta are also shown in Fig.~\ref{3d_timeframe_rho_difference}.

%\subsection{Closer to the boundaries}

The simulations were run under the hypothesis that the interaction region is smaller than the jet radius, $R_{\rm j}$, and much smaller than the jet scale height, allowing us to assume that the ejecta is immersed in a homogeneous flow.
However, the scales reached by the shock, in particular in the larger grid S2 simulations, indicate that the shocked ejecta could easily reach the boundaries of the jet and strongly interact with the shear layer and the interstellar medium. This phenomenon could trigger additional shocks, and thus turbulence and mixing between the jet flow, the ejecta material and the interstellar medium gas. We will explore this scenario in upcoming work. Moreover, the presence of magnetic fields can strongly impact the dynamical and non-thermal processes involved in the jet-ejecta interaction. This possibility will also be explored by means of relativistic magnetohydrodynamical simulations, using the code \texttt{LOSTREGO} \citep{miralles22}.

%\subsection{Non-thermal emission}

In terms of detectability, part of the jet power is reprocessed at the shock for most of the interaction duration, that is, as long as the ejecta has not been accelerated to the jet bulk speed. Therefore, plenty of energy is likely to feed non-thermal processes. Synchrotron and inverse Compton (IC) in the case of electrons would in principle be more efficient than hadronic processes, given the large scale of the interaction and diluted target fields, although specific calculations are required for a quantitative assessment. Future work will be devoted to the non-thermal consequences of the jet-ejecta interaction. Nevertheless, it is worth indicating that, unless strongly below equipartition, the shocked jet and ejecta magnetic fields make synchrotron emission strongly dominant over adiabatic or IC losses, releasing most of the energy of non-thermal electrons (and positrons) from radio to X-rays. Moreover, the spatial scales in which the jet-ejecta interaction can be more violent, $\sim 100$~pc, translate into angular sizes of a few tens of milliarcseconds for an AGN at Gpc distances. This makes the scenario potentially resolvable by radio interferometric arrays, like VLA in configuration A at $\nu \geq 22$~GHz and certainly for any VLBI array, and also by Hubble in nearby sources, or Chandra for the nearest ones. 
If hadrons were also accelerated, efficient neutrino production could occur either from proton-proton (p-p) or proton-photon (p-$\gamma$) interactions, in particular for a more powerful jet with an interaction region closer to its base. Given the relatively high density of mass-loaded protons, p-p interactions could be favored here \citep[as suggested in][in the case of TXS\,0506+056 for a jet-cloud/star interaction]{2020A&A...633L...1R,wang2022}, producing TeV to PeV neutrinos. Protons could reach even higher energies and interact with intense photon fields allowing the production of pairs, and more energetic neutrinos (> PeV) trough p-$\gamma$ interactions. Futures studies will help us to assess the detectability of such an event with current and future neutrino facilities. Finally, the simulations show that jet-ejecta mixing is very efficient, so heavy nuclei entrained by the jet could easily reach the jet shock or the regions with strong velocity shear and turbulence downstream of the interaction, having there a high chance of getting accelerated up to ultra-high energies, as proposed in \cite{bosch23}.

\begin{acknowledgements}
This work has received financial support from the Spanish Ministry of Science and Innovation under grants 
PID2022-136828NB-C41/AEI/10.13039/501100011033/ERDF/EU and PID2022-136828NB-C43/AEI/10.13039/501100011033/ERDF/EU, through the María de Maeztu 2020-2023 award to the ICCUB (CEX2019-000918-M), and from the Generalitat de Catalunya through grant 2021SGR00679. V.B-R. is Correspondent Researcher of CONICET, Argentina, at the IAR. We thank the anonymous referee for a constructive report that has improved the paper.
\end{acknowledgements}

\bibliographystyle{aa}
\bibliography{biblio}

\appendix
\section{Simulations table}

\begin{table}
\caption{Simulation parameters; common parameters at the bottom.}             % title of Table
\label{table:1}      % is used to refer this table in the text
\centering                          % used for centering table
\begin{tabular}{c c c c c c c c c}        % centered columns (4 columns)
\hline
%
%%%%%%%%%%%%%%%%%%%%%%%%%%%%%%%%%%%%%%%%%%%%%%%%%%%%%%%%%%%%%

Simulation &  $R_{\rm SN} (\rm pc)$ & $L_x\,(R_{\rm SN})$ & $L_y\,(R_{\rm SN})$ & $L_z\,(R_{\rm SN})$ & SN location (jet axis) & $\rho_{\rm SN}\,(\text{g}{\,}\text{cm}^{-3})$ & SN motion $v_{\rm orb}^{x}$ & axisymmetry   \\    % table heading 
                
   S1 2D & 1.1 &$96$ & - & $192$ & $L_z/4$ & $2.4\times10^{-23}$ & no & yes \\    

   S1 3D & 1.1  &$100$ & $100$ & $100$ & $L_y/5$ & $2.4\times10^{-23}$ & yes & no \\
                                 
   S2 2D & 2.2 &$80$ & - & $120$ & $L_z/4$ & $3\times10^{-24}$ & no & yes \\
                            
   S2 3D & 2.2 &$80$ & $80$ & $80$ & $L_y/8$ & $3\times10^{-24}$ & yes & no\\
\hline       
\vspace{.1cm}
$R_{\rm SN}$ (cells) &  $L_{\rm j}\,(\rm erg\,s^{-1})$ & $R_{\rm j}\,(\rm pc)$ & $W_{\rm j}$ & $E_{\rm SN}\,(\rm erg)$ & $M_{\rm SN}\,(M_\odot)$ & $\rho_{\rm j}\,(\rm g\,\rm cm^{-3})$ & $T_{\rm j}\,(K)$ & $T_{\rm SN}\,(K)$ \\          
8 & $10^{44}$ & $100$ & $2$ & $10^{51} $ & $2$ & $6\times10^{-30}$ & $2\times10^{11} $ & $10^{9}$ \\
\hline             
\end{tabular}
\end{table} 
\section{3D simulations with a supernova remnant at rest}
\label{3dwm}
%%%%%%%%%%%%%%%%%%%%%%%%%%%%%%%%%%%%%%%%%%%%%%%%%%%%%
%

In this Appendix, we show the results of 3D simulations of the jet/SN remnant interaction, where the SN progenitor is initially at rest. With all the remaining numerical and physical characteristics being equal to those of simulations S1 and S2 discussed in the main body of this work, these initially symmetric simulations allow us to assess the effects of the orbital speed of the progenitor star around the galactic center consider in the original simulations.

\begin{figure*}[]
\centering
\includegraphics[width=.95\linewidth]{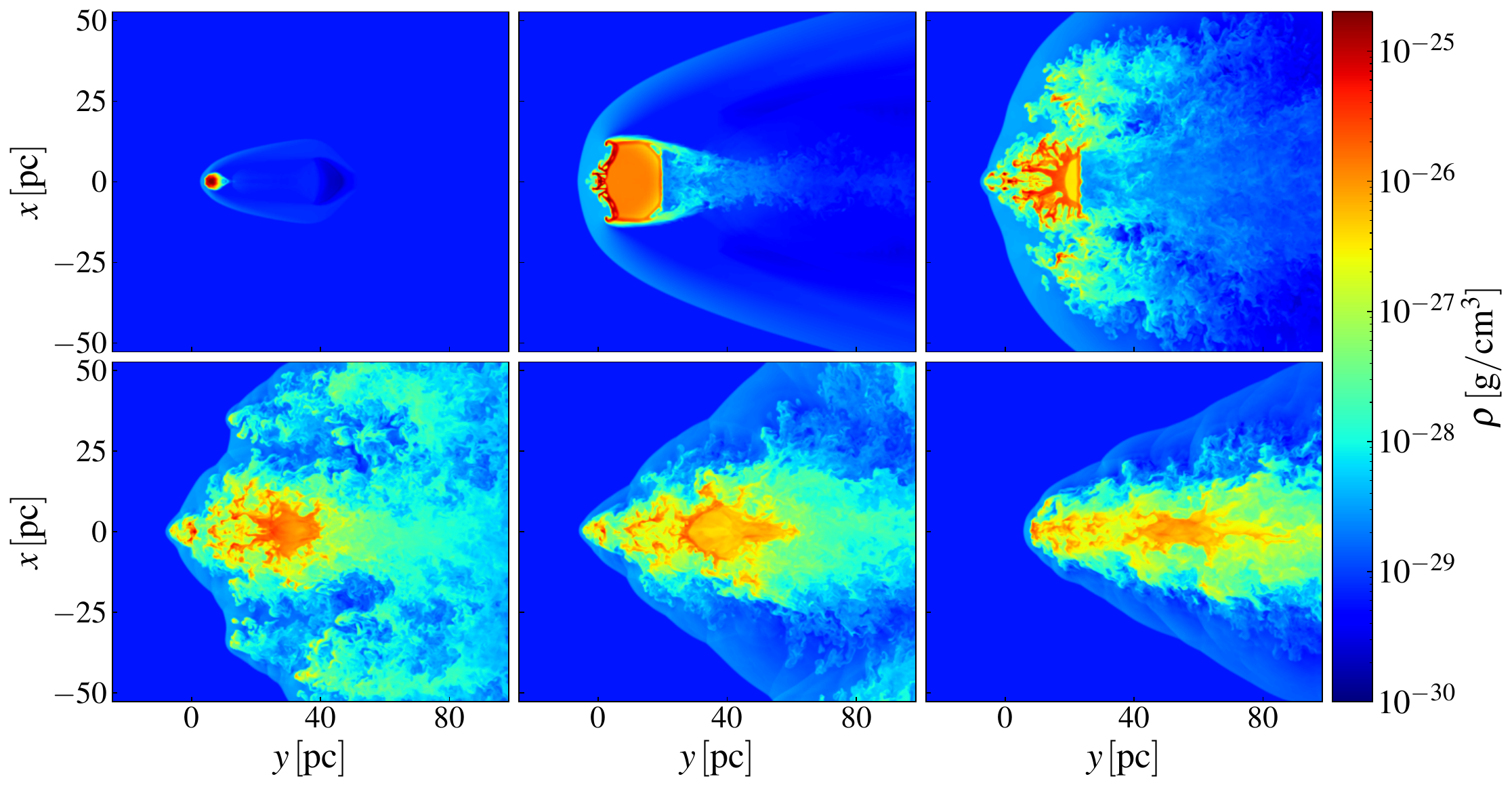}
\caption{2D cuts of the 3D S1 simulation showing density in the $XY$-plane at $z=0$ with an initially static ejecta. Top left panel: $t\approx140\,$yr; top middle panel: $t\approx1600\,$yr; top right panel: $t\approx2400\,$yr; bottom left panel: $t\approx3200\,$yr; bottom middle panel: $t\approx3700\,$yr; bottom right panel: $t\approx4900\,$yr.}
\label{3d_timeframe_rho_hr_wm}
\end{figure*}

\begin{figure*}[]
\centering
\includegraphics[width=.94\linewidth]{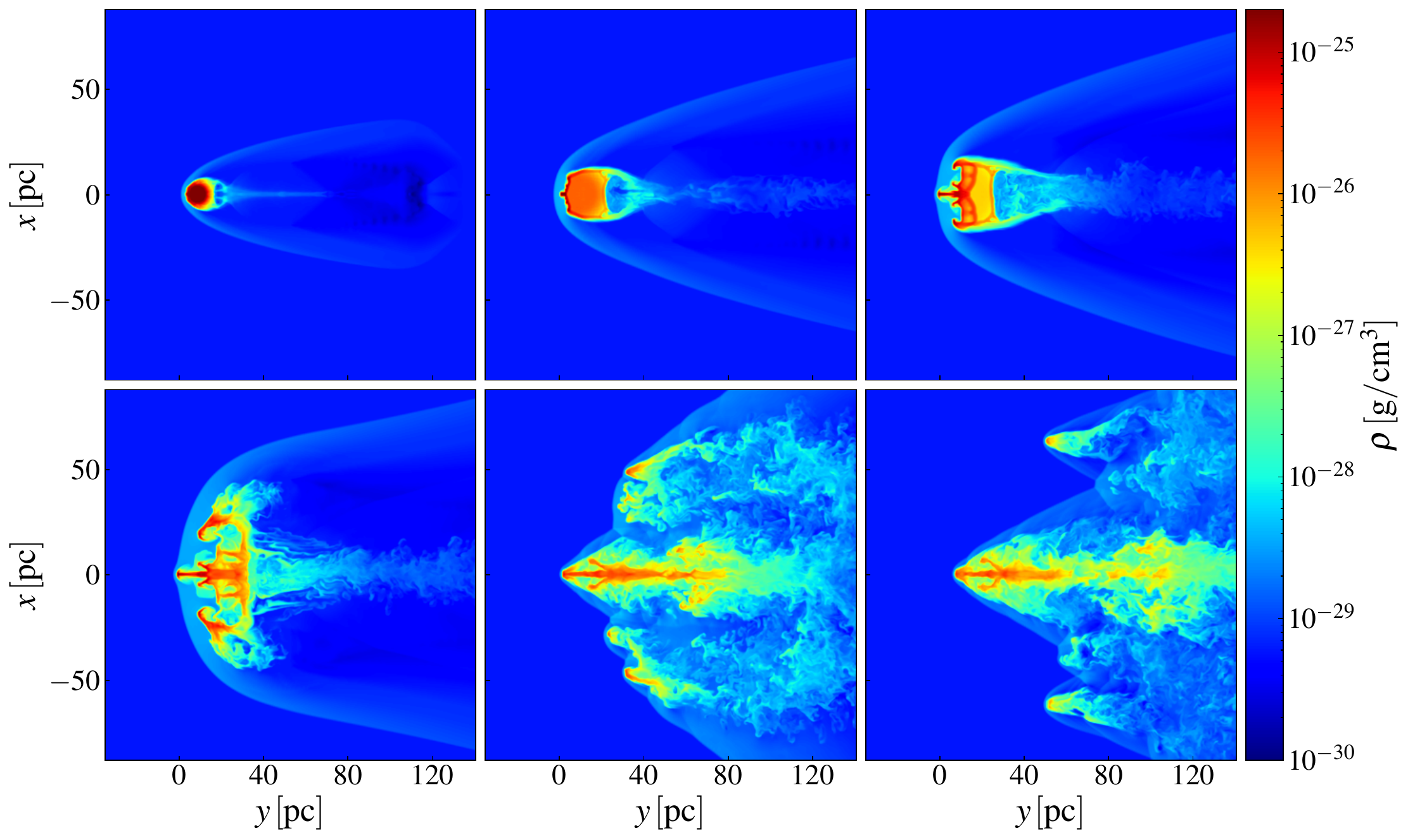}
\caption{2D cuts of the 3D S2 simulation showing rest-mass density in the $XY$-plane at $z=0$ with an initially static ejecta. Top left panel: $t\approx430\,$yr; top middle panel: $t\approx1400\,$yr; top right panel: $t\approx2300\,$yr; bottom left panel: $t\approx2900\,$yr; bottom middle panel: $t\approx4000\,$yr; bottom right panel: $t\approx4700\,$yr.}
\label{3d_timeframe_rho_hl_wm}
\end{figure*}

\begin{figure*}[]
\centering
\includegraphics[width=.94\linewidth]{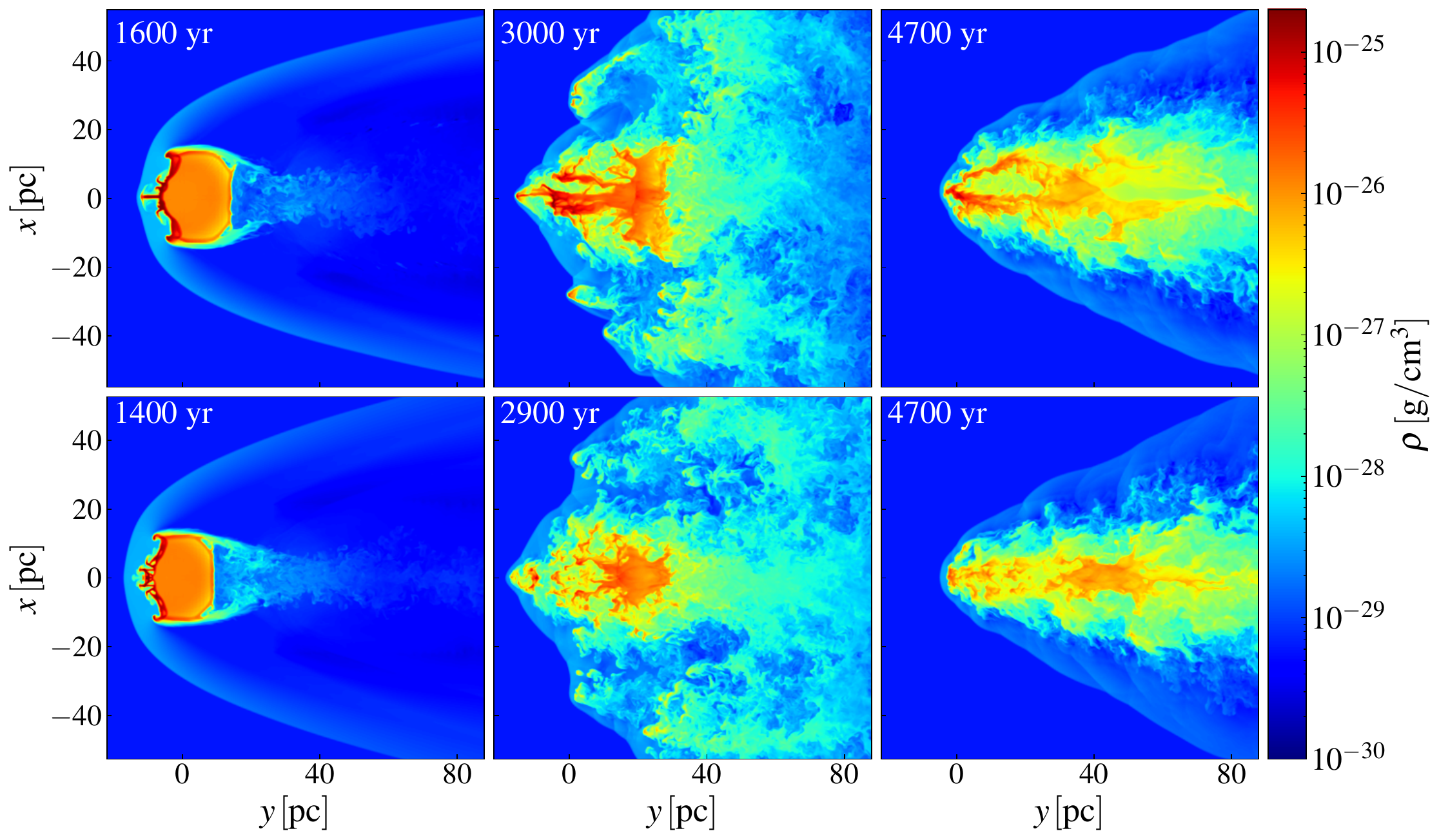}
\caption{2D cuts of the 3D S1 simulation showing rest-mass density in the $XY$-plane at $z=0$, with (top panels) and without (bottom panels) motion of the ejecta. We selected here 3 different moments from the simulation (e.g. \ref{3d_timeframe_rho_hr_wm} and \ref{3d_timeframe_rho_hr}) to show the asymmetries introduced by the orbital motion $v^{x}_{\rm orb}$.}
\label{3d_timeframe_rho_difference}
\end{figure*}

Figures~\ref{3d_timeframe_rho_hr_wm} and \ref{3d_timeframe_rho_hl_wm} show the rest-mass density in the $XY$-plane at $z=0$ in a series of snapshots comparable to those in Figs.~\ref{3d_timeframe_rho_hr} and \ref{3d_timeframe_rho_lr} for simulations S1 and S2, respectively. The different snapshots display a remarkable symmetry between the upper and lower halves along most of the simulation. Small asymmetries of numerical origin \citep{1999ApJS..122..151A,2005A&A...443..863P} are visible by the end of the simulations once the remnant has been disrupted. In contrast, noticeable asymmetries are observed much earlier in Fig.~\ref{3d_timeframe_rho_hr} (beyond 2000~yr) and Fig.~\ref{3d_timeframe_rho_lr} (beyond 3000~yr). In any case, the gross morphological and dynamical properties of simulations S1 and S2 remain very close to their static counterparts since the orbital speed of the remnant is very small compared to other characteristic speeds in the flow (e.g., light speed). \\
Figure \ref{3d_timeframe_rho_difference} selects 3 snapshots to be compared between the simulations with (e.g. Fig~\ref{3d_timeframe_rho_hr}) and without motion (e.g. Fig~\ref{3d_timeframe_rho_hr_wm}) to show the asymmetries introduced by $v_{\rm orb}^{x}$.

\clearpage

%
%%%%%%%%%%%%%%%%%%%%%%%%%%%%%%%%%%%%%%%%%%%%%%%%%%%%%

%%%%%%%%%%%%%%%%%%%%%%%%%%%%%%%%%%%%%%%%%%%%%%%%%%%%%
%
%
%%%%%%%%%%%%%%%%%%%%%%%%%%%%%%%%%%%%%%%%%%%%%%%%%%%%%%

%%%%%%%%%%%%%%%%%%%%%%%%%%%%%%%%%%%%%%%%%%%%%%%%%%%%%
%
%
%%%%%%%%%%%%%%%%%%%%%%%%%%%%%%%%%%%%%%%%%%%%%%%%%%%%%%

\end{document}